\immediate\write18{makeindex \jobname.nlo -s nomencl.ist -o \jobname.nls}

\documentclass[3p]{elsarticle}

\UseRawInputEncoding
\usepackage{tabularx}
\usepackage{amsmath}
\usepackage{amssymb}
\usepackage{graphicx}
\usepackage{subfigure}
\usepackage{enumerate}
\usepackage{float}
\usepackage[percent]{overpic}
\usepackage{sidecap}
\usepackage{tipa}
\usepackage{epstopdf}
\usepackage{booktabs}
\usepackage{natbib}
\usepackage{hyperref}
\usepackage{lineno}
\usepackage{appendix}
\usepackage{arydshln}
\usepackage{mathtools}
\usepackage{slashbox}
\usepackage{diagbox}

\usepackage{scalerel,stackengine,amsmath}
\newcommand\equalhat{\mathrel{\stackon[1.5pt]{=}{\stretchto{%
    \scalerel*[\widthof{=}]{\wedge}{\rule{1ex}{3ex}}}{0.5ex}}}}

\usepackage{tikz}
\DeclareRobustCommand*\circled[1]{\tikz[baseline=(char.base)]{
\node[shape=circle,draw,inner sep=1pt] (char) {#1};}}

\modulolinenumbers[5]

\DeclareMathOperator{\sign}{sign}

\usepackage{framed} 
\usepackage[nomentbl,stdsubgroups,nocfg]{nomencl} 

\makenomenclature

\makeatletter
\def\ps@pprintTitle{%
  \let\@oddhead\@empty
  \let\@evenhead\@empty
  \let\@oddfoot\@empty
  \let\@evenfoot\@oddfoot
}
\makeatother

\usepackage{numcompress}\biboptions{authoryear}

\begin{document}

\begin{frontmatter}

\title{Cause-effect relationship between model parameters and damping performance of hydraulic shock absorbers}



\author[mainaddress]{Lukas Schickhofer\corref{correspondingauthor}}
\address[mainaddress]{KTM Research \& Development, Mattighofen, Austria}
\cortext[correspondingauthor]{Corresponding author:}
\ead{lukas.schickhofer13@alumni.imperial.ac.uk}

\author[secondaryaddress]{Chris G. Antonopoulos}
\address[secondaryaddress]{School of Mathematics, Statistics and Actuarial Science (SMSAS), University of Essex, Wivenhoe Park, United Kingdom}

\begin{abstract}
Despite long-term research and development of modern shock absorbers, the effect of variations of several crucial material and model parameters still remains dubious. The goal of this work is therefore a study of the changes of shock absorber dynamics with respect to typical parameter ranges in a realistic model. We study the impact of shim properties, as well as geometric features such as discharge coefficients and bleed orifice cross section. We derive cause-effect relationships by nonlinear parameter fitting of the differential equations of the model and show digressive and progressive quadratic damping curves for shim number and thickness, sharp exponential curves for discharge coefficients, and leakage width, as well as a linear decrease of damping properties with bleed orifice area. Temperature increase affecting material properties, such as density and viscosity of the mineral oil, is found to have a mostly linear relationship with damping and pressure losses. Our results are not only significant for the general understanding of shock absorber dynamics, but also serve as a guidance for the development of specific models by following the proposed methodology.
\end{abstract}


\end{frontmatter}

\section{\textbf{Introduction}}
\label{Sec:1}

The development and tuning of shock absorbers leads them to effectively absorb displacements and excitations due to road irregularities. This is particularly important for motorcycles and bikes, where the contact area with the ground is small and wheel grip must not be compromised even in the most extreme situations (e.g. large tilting angles, sharp corners, sudden change in road properties). Hydraulic shock absorbers play a crucial role in the suspension of cars, trains, and even airplanes \citep{boggs2010efficient,wang2017rail}. 
As a result, several types of shock absorbers are applied to a wide range of vehicles, which complicates efforts to find a universal analytical model for their behaviour and to reach general conclusions regarding their sensitivity to parameter and model changes.
The most common type of shock absorber in use today is the \emph{monotube damper}, where a piston carrying stacks of metal plates (i.e. \emph{shims}) separates two chambers filled with oil, which are adjacent to a nitrogen compartment to absorb large volumetric changes. Further developments of this basic setup are e.g. the \emph{piggyback damper}, or the \emph{dual-tube damper} \citep{dixon2008shock}.

Due to their complex and intricate setup that involves several multi-component valves and chambers filled with viscous fluid, hydraulic shock absorbers have always been difficult to develop and adjust to the specific requirements of road vehicles. This is the main reason why a large-scale analysis of the dynamical behaviour of shock absorbers within parameter spaces across multiple dimensions and excitation frequencies is important. 

Despite this obvious need, only few studies exist on the sensitivity of damping performance to parametric changes:
\citet{talbott2002experimentally} derived a linearized, algebraic model of a monotube shock absorber and performed first investigations of the impact of relevant quantities such as shim stack stiffness and bleed orifice area on the damping characteristics. They were able to correctly predict basic trends of these variables, but did not capture nonlinear effects such as hysteresis or instabilities at high flow rate. Moreover, their shim stack model was rudimentary and did not resolve the force balance across shims and the influence of shim number or thickness.
\citet{farjoud2012nonlinear} formulated a fully nonlinear reduced-order model of a hydraulic damper and solved the elastic valve deformation via the Rayleigh-Ritz method.
The parameter study was limited to the impact of shim disk thickness, stiffness, pre-deflection, and bypass orifice opening.
\citet{skavckauskas2017development} included an explicit computation of shim contacts and deflection in their nonlinear damper model by application of the force method. However, they limited their study to only a few parameters such as cross-sectional orifice area and shim thickness. Additionally, they considered only the direct impact of parameter variations on damping force without considering other crucial shock absorber properties, such as localized hydraulic pressure losses.
Additionally, several authors investigated the application of shock absorber models based on differential equation systems for nonlinear effects, such as fluid compressibility, hysteresis, semi-active and elastic valve displacement, structure-borne noise, as well as failure analysis \citep{duym1997physical,pelosi2013investigation,czop2009simplified,benaziz2015shock,guan2019test,guan2022theoretical}.

Given the consideration of only a limited set of semi-empirical parameters in the aforementioned studies, our work is the first attempt to attain a database of all important shock absorber variables, such as pressure loss and damping force, for vast parameter spaces of a validated physical model. The goal is not only to capture transient effects at high stroking frequencies and varied model dimensions, but also to derive the precise cause-effect relationships between parameter changes and damping characteristics.

\section{\textbf{Methods}}
\label{Sec:2}

In the following, the basic modelling assumptions for this study are outlined, which are described in detail in \citet{schickhofer2023universal}. Figure \ref{Fig:2-1}(a)-(b) outlines the flow directions in and out of the adjacent control volumes of the main piston during engagement of the shock absorber. For the complete damping system, there are several control volumes (i.e. compression and rebound chamber, gas reservoir) that depend on the particular type.
Additionally, Fig. \ref{Fig:2-1}(b) shows the dominant forces acting on a singular shock absorber valve, namely pretension force, pressure force, momentum force, and impact force. The valve acts as a driven oscillator and has a characteristic spring constant and damping coefficient depending on the structural and fluid properties.

\begin{figure}[htbp]
\begin{centering}
\begin{overpic}[width=0.80\textwidth]{Figure1.pdf}
\end{overpic}
\par
\end{centering}
\caption{Sketch of a monotube damper as investigated in this study with all relevant modelling parameters shown, including its rebound chamber \circled{1}, compression chamber \circled{2} (a), as well as flow rates, pressures, and force balance across the shim valves (b)-(c).
The typical monotube shock absorber geometry is depicted by cross-sections (a)-(b), while the reduced diagram (c) depicts the dominant effects acting on the valve, which is modelled as a forced, damped oscillator.}
\label{Fig:2-1} 
\end{figure}

The underlying mathematical model for a shock absorber valve consists of the system of ordinary differential equations
\begin{align}
m\ddot{x}+c\dot{x}+kx &= F_{p}\left(p\right)+F_{m}\left(x,p\right)+F_{i}\left(x\right)-F_{0},  \label{Eq:2A-1}  \\
\dot{p} &= \frac{1}{\beta V}\left[Q-Q_{v}\left(x,p\right)-Q_{b}\left(p\right)\right],  \label{Eq:2A-2} 
\end{align}
which reflect the Newtonian force balance on the valve in Eq. (\ref{Eq:2A-1}) and the mass conservation across the control volume $V$ in Eq. (\ref{Eq:2A-2}).
The main system variables are the valve displacement $x$ and the pressure $p$, which depend on each other.
The applied definitions of pressure force $F_{p}$, momentum force $F_{m}$, impact force $F_{i}$, as well as of valve flow rate $Q_{v}$ and bleed flow rate $Q_{b}$ are given by
\begin{equation}
F_{p}\left(p\right)=A_{p}\Delta p=A_{p}\left(p-p_{0}\right) \equalhat A_{p}p, \nonumber
\label{Eq:2A-3} 
\end{equation}
\begin{equation}
F_{m}\left(x,p\right)=\frac{C_{f}\rho}{A_{p}}\left(\underbrace{\alpha\pi d_{v}x}_{A_{v}\left(x\right)}C_{d,v}\sqrt{\frac{2p}{\rho}}\right)^2, \nonumber
\label{Eq:2A-4} 
\end{equation}
\begin{equation}
F_{i}\left(x\right)=\begin{cases}
-k_{i}\left(x-x_{u}\right), & \quad x\leq x_{u},\\
0, & \quad x_{d}<x<x_{u},\\
-k_{i}\left(x-x_{d}\right), & \quad x\geq x_{d},
\end{cases} \nonumber
\label{Eq:2A-5} 
\end{equation}
\begin{equation}
Q_{v}\left(x,p\right)=C_{d,v}\underbrace{\alpha \pi d_{v}x}_{A_{v}\left(x\right)}\sqrt{\frac{2\Delta p}{\rho}}, \nonumber
\label{Eq:2A-6} 
\end{equation}
\begin{equation}
Q_{b}\left(p\right)=C_{d,b}A_{b}\sqrt{\frac{2\Delta p}{\rho}}. \nonumber
\label{Eq:2A-7} 
\end{equation}
Here, the momentum coefficient $C_{f}\approx0.3$ gives an empirically determined measure of momentum transfer by the jet flow exiting the piston orifices and was first evaluated by \citet{lang1977study} for shock absorber fluid dynamics. 
Moreover, the discharge coefficient $C_{d}=A_{f}/A_{0}$ defines the ratio between the actual flow cross section $A_{f}$ and orifice cross section $A_{0}$. We further distinguish between a valve discharge coefficient $C_{d,v}$, which is related to the shim valve flow $Q_{v}$, and a bleed orifice discharge coefficient $C_{d,b}$ connected to the bleed flow $Q_{b}$. These different flow rates are indicated in Fig. \ref{Fig:2-1}(b). 
The various discharge coefficients of the model, which are further defined in \ref{App:A}, are also a way of taking into account second-order, three-dimensional geometry effects of the surrounding channels and gaskets that would otherwise not be resolved in the reduced-order physical model. As such, they are obtained from validation studies and test bench measurements of the monotube shock absorber geometry \citep{schickhofer2023universal}. A comprehensive overview of discharge coefficients for internal flows in shock absorbers is also given by \citet{dixon2008shock}.
Furthermore, the effective valve flow cross-section $A_{v}\left(x\right)$ is a function of the displacement $x$, shim plate diameter $d_v$, and the effective flow area coefficient $\alpha$, which gives the ratio of valve orifice area to total piston area and depends on the geometry of the applied piston.

Typically, the valve consists of an elastic shim stack, in which case the deflection and stiffness can be computed by a statically indeterminate system using the method of superposition:
By using the partial displacements $\delta_{k}$ of all $n$ shims, the effective total stiffness of a shim stack is determined as
\begin{equation}
k \equalhat k_{tot} = \frac{F_{tot}}{\delta},   \nonumber
\label{Eq:2-1}  
\end{equation}
where $F_{tot}$ is the total force acting on the shim stack and $\delta$ is the resulting displacement of the shim stack equivalent to the displacement $\delta_{n}$ of the last shim.
The vector of single shim displacements $\delta_{k}$ is obtained by 
\begin{equation}
\hat{K}\cdot\vec{w}=\vec{\delta},  \nonumber
\label{Eq:2-2}  
\end{equation}
with the stiffness matrix
\begin{align}
  \hat{K} &=
  \left[ {\begin{array}{cccc}
    c_{\delta_{1}w_{1}} & 0 & 0 & 0  \\
    c_{\delta_{2}w_{1}} & c_{\delta_{2}w_{2}} & 0 & 0  \\
    0 &  \ddots & \ddots & 0 \\
   0 & 0 & c_{\delta_{n-1}w_{n-2}} & c_{\delta_{n-1}w_{n-1}} \\
  \end{array} } \right]  \nonumber
\end{align}
and the vector of reaction forces
\begin{align}
 \vec{w} &= 
  \left[ {\begin{array}{c}
    w_{1}  \\
    w_{2}  \\
    \vdots \\
    w_{n-1} \\
  \end{array} } \right].  \nonumber  \label{Eq:2-4}
\end{align}
Here, the coefficients 
\begin{equation}
c_{\delta_{k}w_{l}} = - \frac{a_{k}}{D} \left[\frac{C_{2}}{C_{8}}\left(\frac{r_{0}C_{9}}{a_{c}}-L_{9}\right)-\frac{r_{0}C_{3}}{a_{c}}+L_{3}\right].  \nonumber
\label{Eq:2-5}
\end{equation}
give the linear displacement $\delta_{k}$ due to the line load $w_{l}$ as in $\delta_{k}=c_{\delta_{k}w_{l}} \cdot w_{l}$ and are composed of the shim plate radii $a_{k}$, the clamping radius $a_{c}$, the radius of acting force $r_0$, the flexural rigidity $D=\left(Et_{k}^{3}\right)/\left[12\left(1-\nu^{2}\right)\right]$ as a function of the shim thicknesses $t_{k}$, and the integration coefficients $C_{2}$, $C_{3}$, $C_{8}$, $C_{9}$, $L_{3}$, $L_{9}$ following from the analytical solution for the deflection of a circular elastic plate by \citet{roark1989formulas}.

Temperature-dependent properties such as viscosity and density of the fluid phase are accounted for in the mathematical model: The viscosity is given by the Guzmann-Carrancio equation,
\begin{equation}
\mu\left(T\right) \approx \mu_{0} e^{C \left(\frac{1}{T}-\frac{1}{T_{0}}\right)},  \nonumber 
\label{Eq:2-6}  
\end{equation}
where a reasonable estimate for the temperature sensitivity coefficient $C$ of regular unimproved mineral oils for dampers is
\begin{equation}
C = 5693 - 304\log_{10}\left(\mu_{15}\right) - 646\log_{10}^{2}\left(\mu_{15}\right).  \nonumber
\label{Eq:2-7}  
\end{equation}
and $\mu_{15}$ is the measured viscosity at \SI{15}{\celsius}, applicable for a range of $0.003 < \mu_{15} < \SI{0.300}{\pascal\second}$ \citep{dixon2008shock}.
The fluid density is modelled as
\begin{equation}
\rho\left(T\right) = \rho_{0} \frac{1}{1+\alpha_{T}\left(T-T_{0}\right)},   \nonumber
\label{Eq:2-8}  
\end{equation}
with the coefficient of volumetric thermal expansion $\alpha_{T} \approx \SI{0.001}{\per\kelvin}$ and the density $\rho_{0} = \rho\left(T=\SI{288}{\kelvin}\right) = \SI{830}{\kilogram\per\cubic\meter}$ at reference temperature $T=T_{0}$.

Using the above expressions, Eqs. (\ref{Eq:2A-1})-(\ref{Eq:2A-2}) can be rewritten as a first-order system of nonlinear, coupled ordinary differential equations
\begin{align}
\dot{y}_{1} &= y_{2},  \label{Eq:2A-8a}  \\
\dot{y}_{2} &= \frac{1}{m} \left[-cy_{2}-ky_{1}+F_{p}\left(y_{3}\right)+F_{m}\left(y_{1},y_{3}\right)+F_{i}\left(y_{1}\right)-F_{0}\right],  \label{Eq:2A-8b}  \\
\dot{y}_{3} &= \frac{1}{\beta V} \left[Q-Q_{v}\left(y_{1},y_{3}\right)-Q_{b}\left(y_{3}\right)\right],  \label{Eq:2A-8c} 
\end{align}
with variables $\vec{y}=\left(y_{1},y_{2},y_{3}\right)=\left(x,\dot{x},p\right)$.

By expanding Eqs. (\ref{Eq:2A-8a})-(\ref{Eq:2A-8c}) for a full monotube shock absorber with its pressure gradients in each chamber and its valve displacements for each shim stack, one obtains the following system:
\begin{align}
\dot{y}_{1} &= y_{4},  \label{Eq:2C-1}  \\
\dot{y}_{2} &= y_{5},  \label{Eq:2C-2}  \\
\dot{y}_{3} &= y_{6},  \label{Eq:2C-3}  \\
\dot{y}_{4} &= \frac{1}{m_{1}} \left( - c_{1}y_{5} - k_{1}y_{1} + F_{p,1} + F_{m,1} + F_{i,1} - F_{0,1} + m_{1}\ddot{x} \right),  \label{Eq:2C-4}  \\
\dot{y}_{5} &= \frac{1}{m_{2}} \left( - c_{2}y_{6} - k_{2}y_{2} + F_{p,2} + F_{m,2} + F_{i,2} - F_{0,2} - m_{2}\ddot{x} \right),  \label{Eq:2C-5}  \\
\dot{y}_{6} &= \frac{1}{m_{g}} \left[ A_{g}\left(y_{7}-y_{9}\right) - F_{f,g} \right],  \label{Eq:2C-6}  \\
\dot{y}_{7} &= \frac{1}{\beta \left[A_{2}\left(L_{2}-x\right)+A_{g}y_{3}\right]} \left( - Q_{v,1} - Q_{v,2} - Q_{b} - Q_{l} + A_{2}\dot{x} - A_{g}y_{6} \right),  \label{Eq:2C-7}  \\
\dot{y}_{8} &= \frac{1}{\beta \left[A_{1}\left(L_{1}+x\right)\right]} \left( Q_{v,1} + Q_{v,2} + Q_{b} + Q_{l} - A_{1}\dot{x} \right),  \label{Eq:2C-8}  \\
\dot{y}_{9} &= - \gamma \frac{y_{9}\left(-A_{g}y_{6}\right)}{A_{g}\left(L_{g}-y_{3}\right)}.  \label{Eq:2C-9} 
\end{align}
The resulting system variables are
\begin{align}
    \vec{y} &= \begin{bmatrix}
              y_{1} \\
              y_{2} \\
              y_{3} \\
              y_{4} \\
              y_{5} \\
              y_{6} \\
              y_{7} \\
              y_{8} \\             
              y_{9}                 
             \end{bmatrix} =
              \begin{bmatrix}
              x_{1} \\
              x_{2} \\
              x_{g} \\
              \dot{x}_{1} \\
              \dot{x}_{2} \\
              \dot{x}_{g} \\
              p_{2} \\
              p_{1} \\             
              p_{g}                 
             \end{bmatrix} \equiv
             \begin{bmatrix}
              x_{c} \\
              x_{r} \\
              x_{g} \\
              \dot{x}_{c} \\
              \dot{x}_{r} \\
              \dot{x}_{g} \\
              p_{c} \\
              p_{r} \\             
              p_{g}                 
             \end{bmatrix}, \nonumber
\label{Eq:2C-10}             
\end{align}
where the symbol $\equiv$ expresses the equivalence between the numbered variables in Fig. \ref{Fig:2-1} and the compression quantities denoted by $c$ and the rebound quantities denoted by $r$, respectively. These are precisely the pressures and valve displacements (and their gradients) associated with the compression and rebound stroke of the shock absorber.

By connecting the analytical expressions for pressure gradients, valve deflections, and rod displacement due to road excitations, a system of first-order, coupled, nonlinear equations is obtained (see Eqs. (\ref{Eq:2C-1})-(\ref{Eq:2C-9})). Furthermore, the analytical model can be generalised in order to describe various shock absorbers with multiple pistons and valves, as well as different architectures. The current version also takes into account nonlinear crossover shim deflection, inertia of the rod, and frictional effects. 
The model defined by Eqs. (\ref{Eq:2C-1})-(\ref{Eq:2C-9}) was also recently validated by \citet{schickhofer2023universal} against the test bench data of a typical shock absorber's damping characteristics.
In these validation studies, we found the modelling results for damping force and pressure drop to be consistently up to 5\% accurate with the experimental datasets.
The system of differential equations is integrated using implicit boundary differentiation formulas (BDF) or hybrid explicit-implicit schemes of Runge-Kutta type (e.g. ADAMS), which are known to be superior for the solutions of \emph{stiff equations}. More importantly, it is possible to perform a detailed investigation of the nonlinear dynamics of the complete system in order to map out its behaviour at sudden, high-frequency excitation and high flow rates.

For this work, a set of parameter values from a realistic shock absorber geometry is applied, which is also listed in \ref{App:A} and serves as a baseline for the subsequent sensitivity studies presented in Sec. \ref{Sec:3}.

\section{\textbf{Results}}
\label{Sec:3}

Crucial parameters of the shock absorber model and their impact on pressure losses and damping force are investigated below. 
The damping force, which is defined as
\begin{equation}
F_{d} = - p_{1}A_{1} + p_{2}A_{2} + \sign \left(p_{1}-p_{2}\right) F_{f} + m_{tot}\ddot{x}, 
\label{Eq:3-1}  
\end{equation}
constitutes the main force exercised by the shock absorber during engagement and is typically given as a function of the rod displacement $x$, rod velocity $\dot{x}$, and the seal friction force $F_{f}$.
Alongside the damping force, the pressure drop across the main piston is given by
\begin{equation}
\Delta p = \left| p_{1}-p_{2} \right|, 
\label{Eq:3-2}  
\end{equation}
and as indicated in Fig. \ref{Fig:2-1}(b), it is an essential feature of a shock absorber and reflects its ability to quickly and effectively dissipate large pressure peaks for a large range of volumetric flow rates through the piston orifices.
The impact of typical shim properties is considered first (cf. Sec. \ref{Sec:3-1}-\ref{Sec:3-3}), before moving on to variables of the piston geometry and fluid properties (cf. Sec. \ref{Sec:3-6}-\ref{Sec:3-5}).

In the sensitivity studies below, variations of particular model parameters from the baseline shock absorber settings given in \ref{App:A} are performed. Consequently, the resulting impact of these changes on the damping behaviour are analysed in the form of parameter spaces and cause-effect diagrams. The fitting functions and their parameters leading to the cause-effect diagrams are confirmed by the $R^2$-value as a statistical measure of the distance of the data points to the fitted regression line. It is also known as the coefficient of determination and goodness-of-fit, with values close to one indicating good suitability of parameters and functions.

\subsection{\textbf{Shim number}}
\label{Sec:3-1}

\begin{figure}[htbp]
\begin{centering}
\begin{overpic}[width=1.00\textwidth]{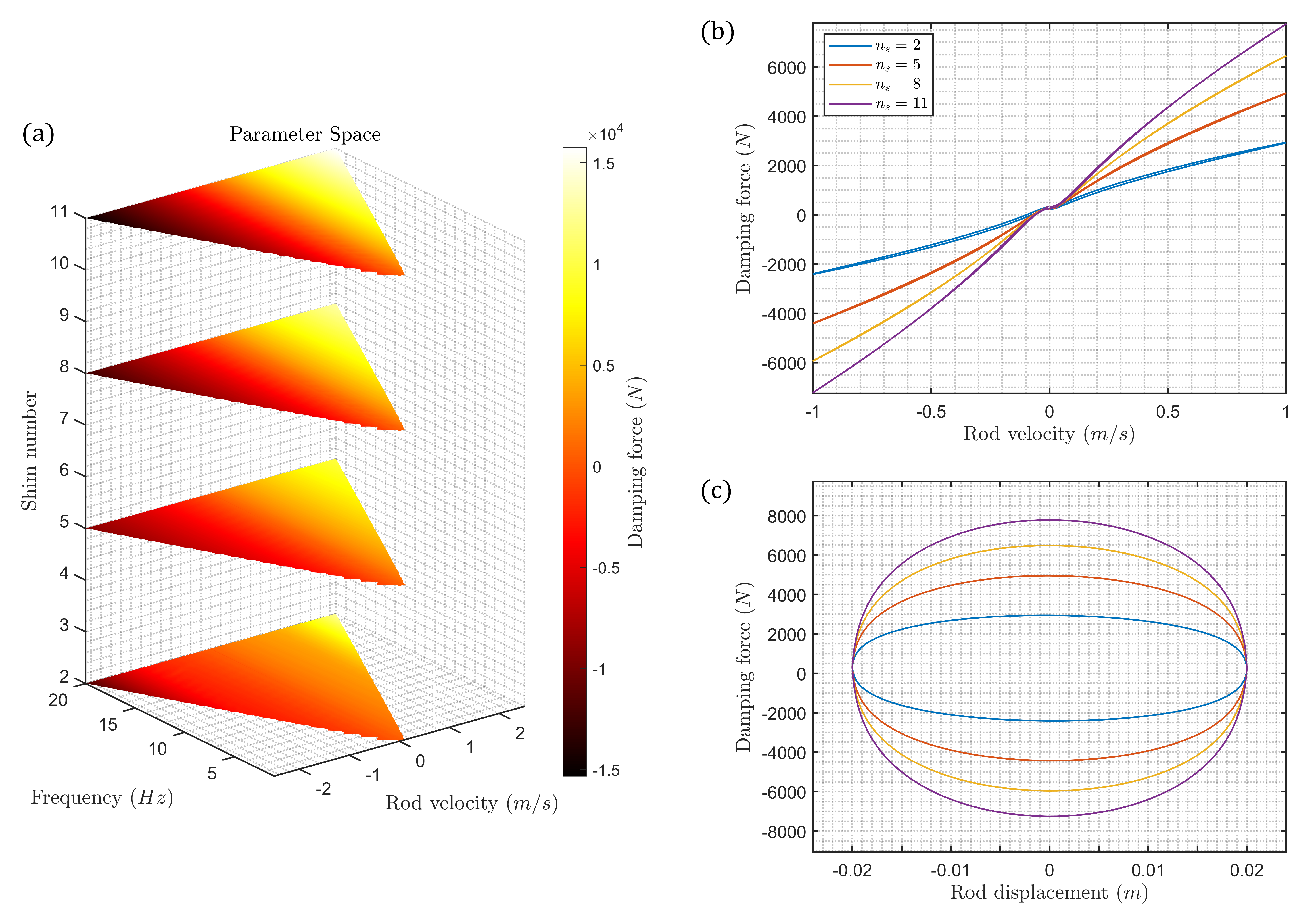}
\end{overpic}
\par
\end{centering}
\caption{Damping force as a function of shim number, frequency, and rod velocity (a). Damping characteristics are also given at a discrete excitation frequency of  $\SI{8}{\hertz}$ for various shim numbers as functions of rod velocity and displacement throughout a full compression-rebound cycle (b),(c).}
\label{Fig:3-1a} 
\end{figure}

\begin{figure}[htbp]
\begin{centering}
\begin{overpic}[width=1.00\textwidth]{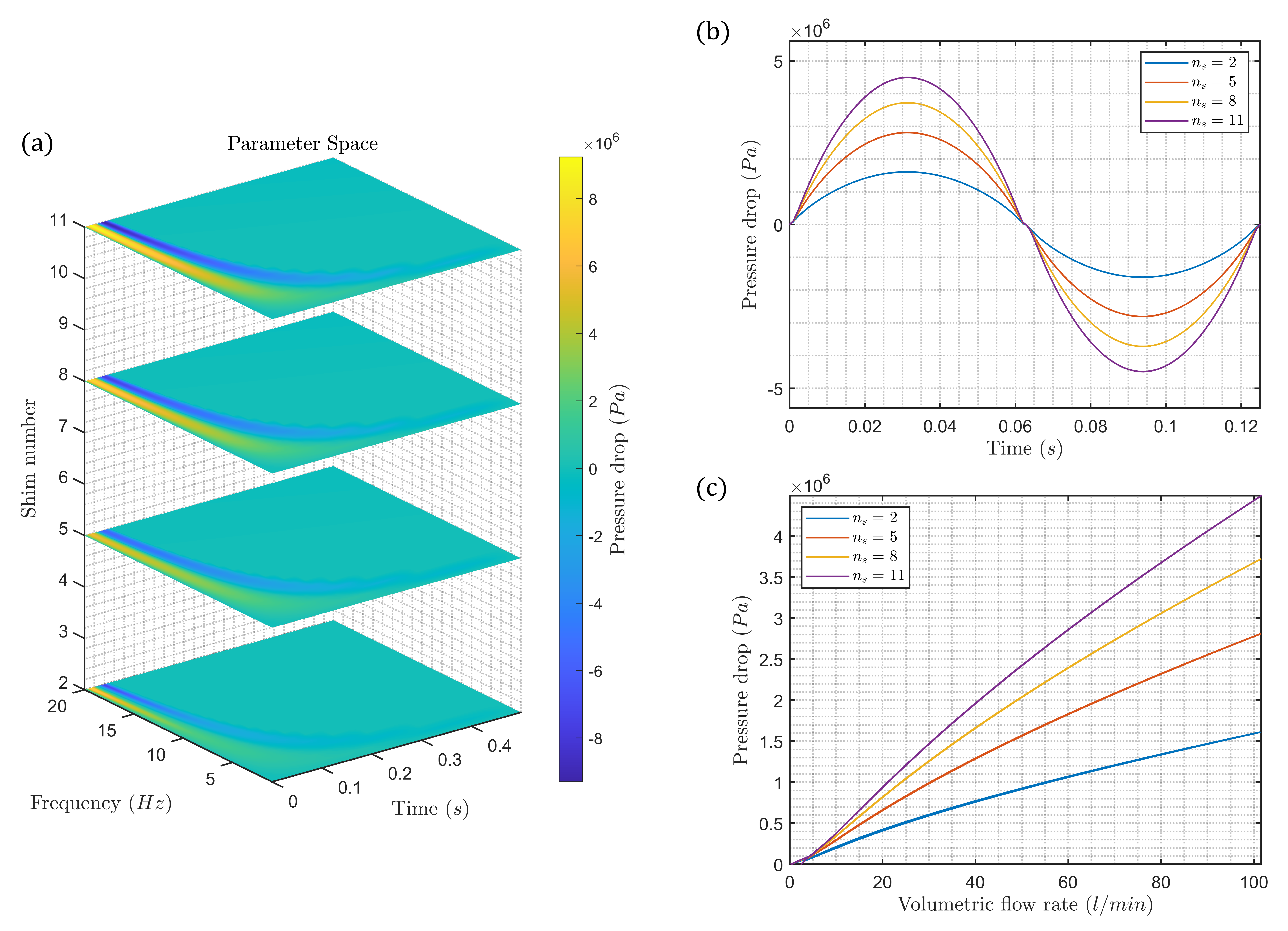}
\end{overpic}
\par
\end{centering}
\caption{Pressure drop across the shock absorber piston as a function of shim number, frequency, and time (a), along side plots of the pressure loss with respect to time and volumetric flow rate across the entire cycle (b),(c).}
\label{Fig:3-1b} 
\end{figure}

\begin{figure}[htbp]
\begin{centering}
\begin{overpic}[width=1.00\textwidth]{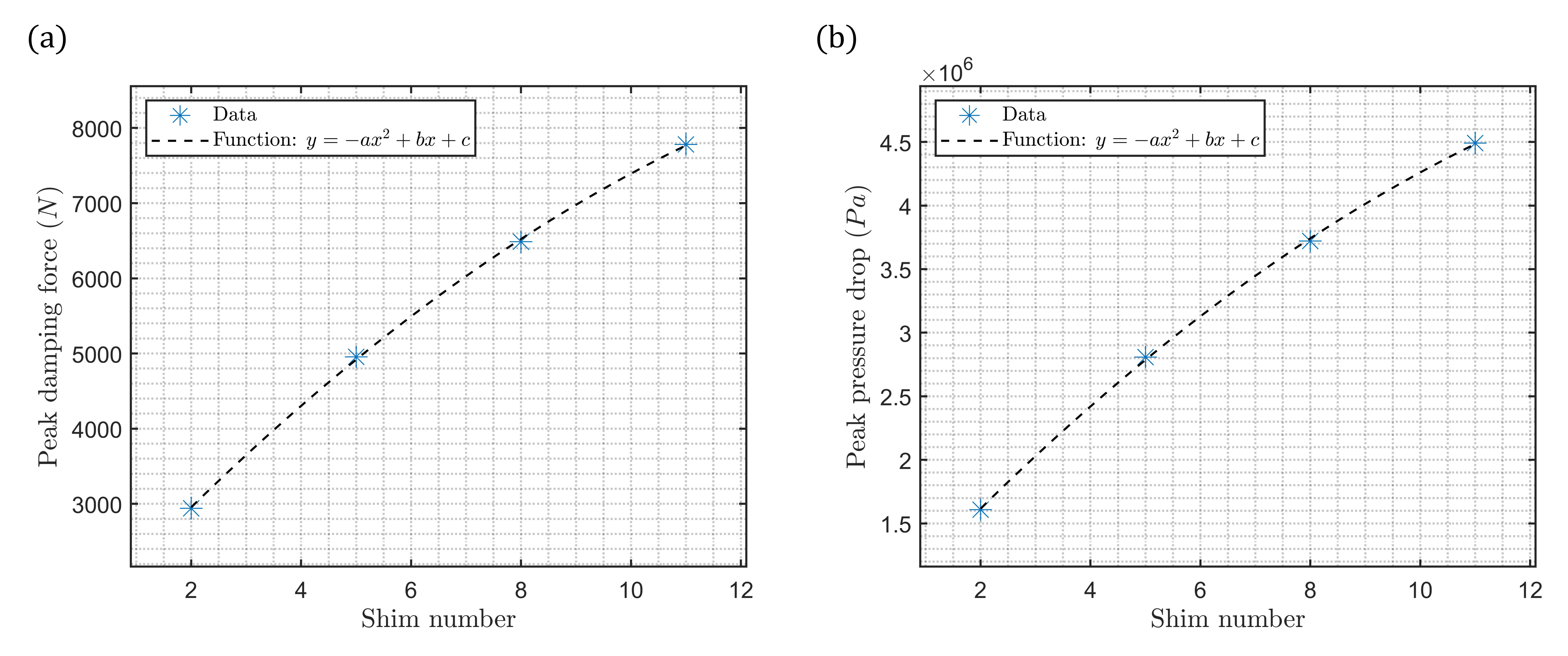}
\end{overpic}
\par
\end{centering}
\caption{Cause-effect relationship of peak damping force (a) and peak pressure drop (b) with respect to changes in the shim number. The fitting parameters are $\left[a,b,c\right]=\left[19.98,794.52,1.45\times10^{3}\right]$ with $R^{2}=0.9986$ in (a) and $\left[a,b,c\right]=\left[1.19\times10^{4},4.73\times10^{5},7.17\times10^{5}\right]$ with $R^{2}=0.9996$ in (b).}
\label{Fig:3-1c} 
\end{figure}

A straightforward way to modify the stiffness of a shim stack is the addition or reduction of the total number of shims of certain predefined thickness. 
By increasing the number of elastic metal shims of a stack (an example of which is pictured in Fig. \ref{Fig:2-1}), a significant rise of overall damping force is reached.
As demonstrated in Fig. \ref{Fig:3-1a}(a), the damping force is also larger for higher frequencies due to the effect of larger rod acceleration and pressure drop within its rebound chamber, causing large gradients across the piston. In Fig. \ref{Fig:3-1b}, it can be seen that the resulting pressure drop, which varies with increasing shim number and stiffness, leads to a higher potential of the shock absorber to dissipate pressure across the considered volumetric flow rates. 
Moreover, as visible in Fig. \ref{Fig:3-1c}, the relationship between changes in the shim number and peak damping force and pressure drop follows a digressive, quadratic curve. 
The data points on this curve are based on the peak damping values for a standard frequency of $f_0=\SI{8}{\hertz}$ on the plane sections through the parameter space of Fig. \ref{Fig:3-1a}(a), which consists of the results of a significant number of simulation runs for each discrete frequency.
The frequency $f_0$ is a typical test excitation frequency for a shock absorber of the investigated type.
Additionally, the resulting parameters of the nonlinear fitting have been computed with the Levenberg-Marquardt algorithm \citep{levenberg1944method, marquardt1963algorithm}.

\subsection{\textbf{Shim thickness}}
\label{Sec:3-2}

\begin{figure}[htbp]
\begin{centering}
\begin{overpic}[width=1.00\textwidth]{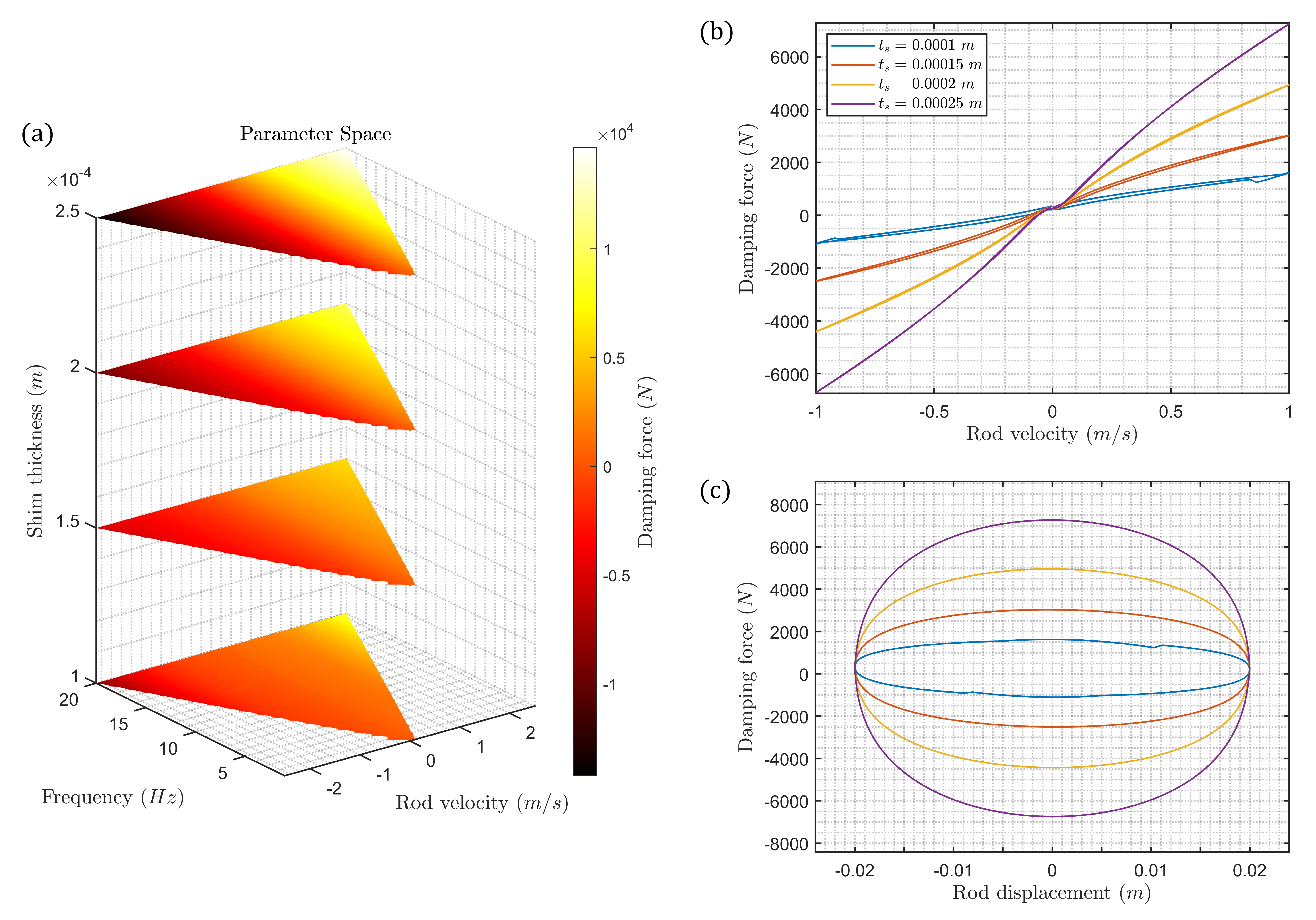}
\end{overpic}
\par
\end{centering}
\caption{Parameter space depicting the dependence of damping force on shim thickness, frequency of excitation, and rod velocity (a). Plots of damping characteristics for different values of the shim thickness at a frequency of $\SI{8}{\hertz}$ are given in (b),(c).}
\label{Fig:3-2a} 
\end{figure}

\begin{figure}[htbp]
\begin{centering}
\begin{overpic}[width=1.00\textwidth]{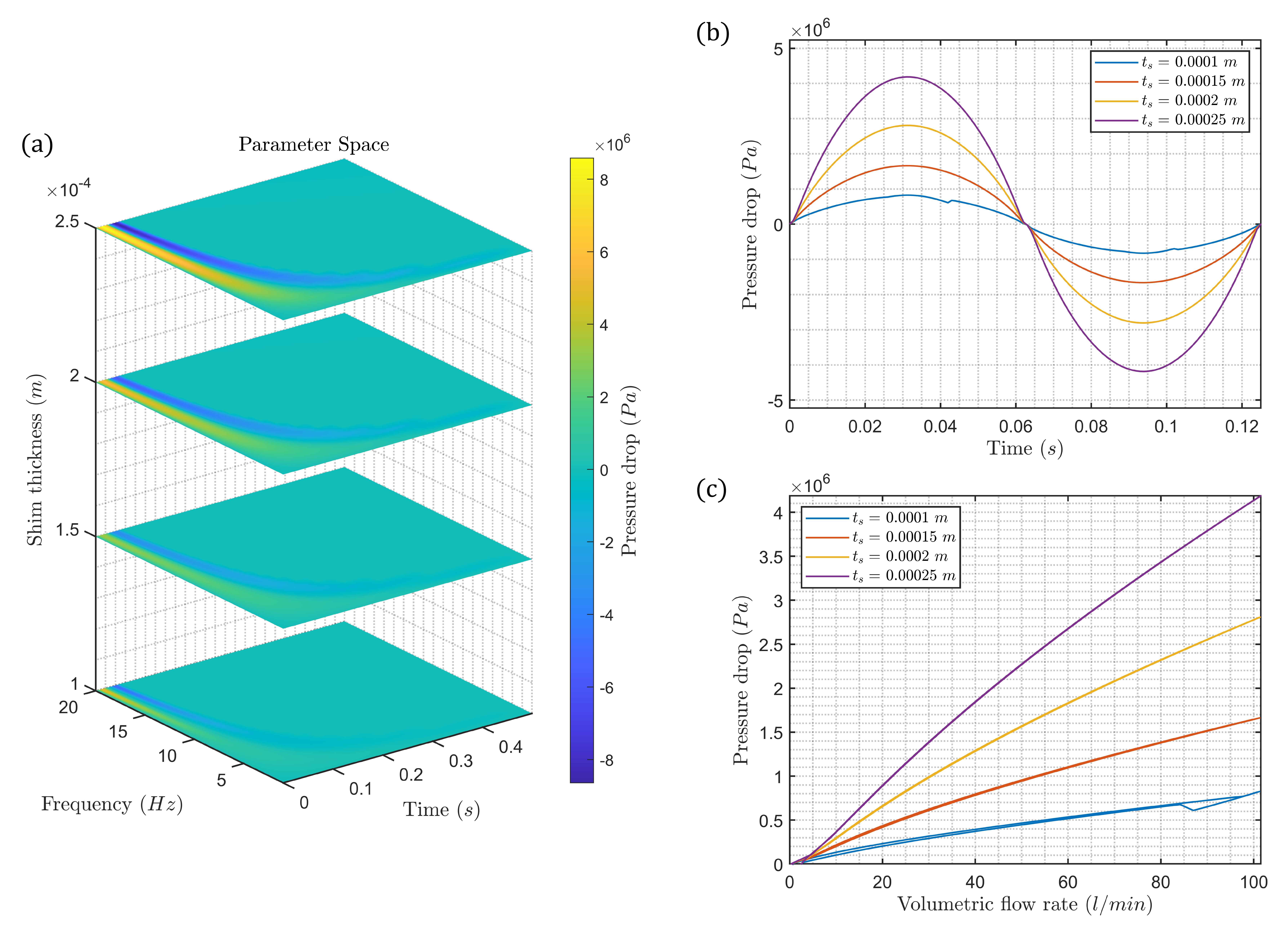}
\end{overpic}
\par
\end{centering}
\caption{Dependence of pressure drop on shim thickness, frequency, and time (a), alongside results of the pressure drop for various shim thicknesses (b) and volumetric flow rates (c).}
\label{Fig:3-2b} 
\end{figure}

\begin{figure}[htbp]
\begin{centering}
\begin{overpic}[width=1.00\textwidth]{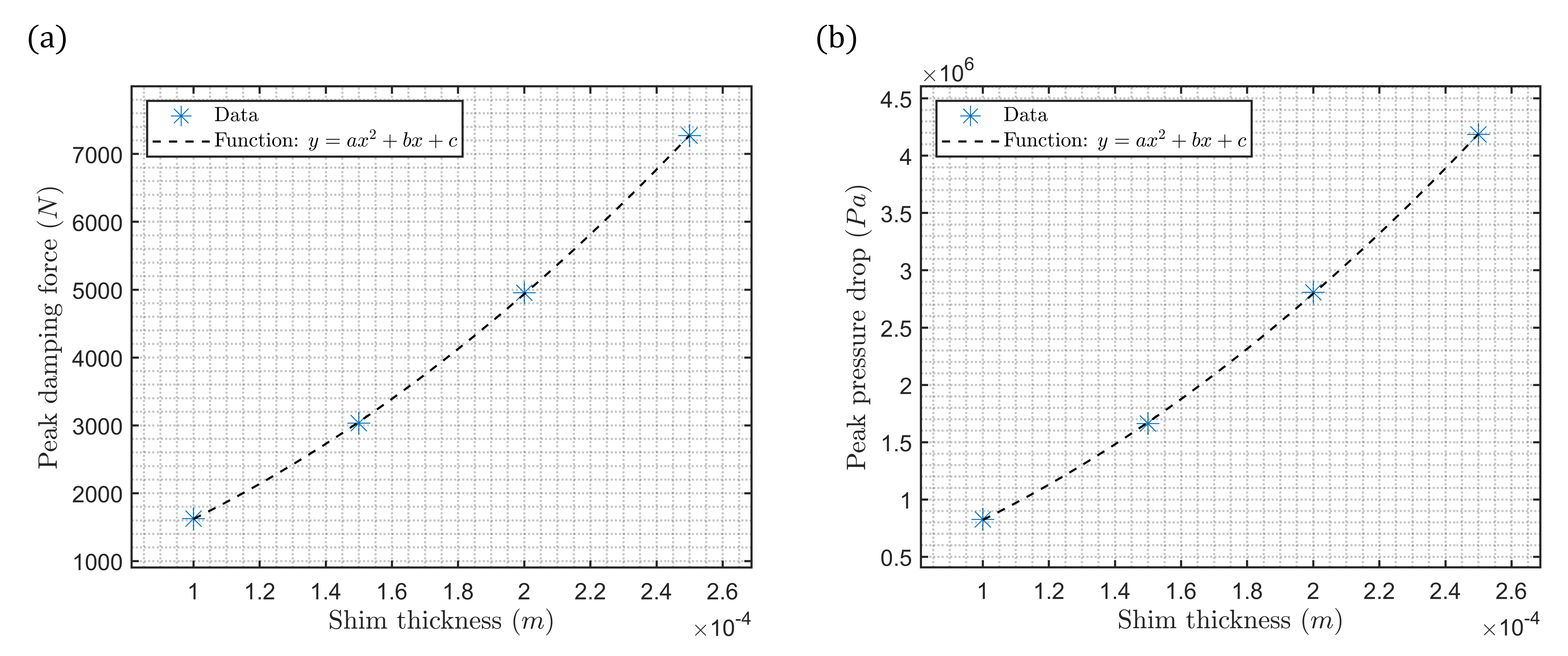}
\end{overpic}
\par
\end{centering}
\caption{Cause-effect relationship between peak damping force (a), or peak pressure drop (b) and shim thickness. Nonlinear fitting gives the parameters $\left[a,b,c\right]=\left[9.10\times10^{10},5.85\times10^{6},125.78\right]$ with $R^{2}=0.9996$ in (a) and $\left[a,b,c\right]=\left[5.42\times10^{13},3.48\times10^{9},-6.89\times10^{4}\right]$ with $R^{2}=0.9999$ in (b).}
\label{Fig:3-2c} 
\end{figure}

Additionally to a change in the shim number (cf. Sec. \ref{Sec:3-1}), the stiffness of a shim stack can be modified by using shims of different thickness. The effect of shim thickness on the damping behaviour is presented in Figs. \ref{Fig:3-2a},\ref{Fig:3-2b}. A crucial difference to the impact of shim number is evident for the cause-effect relationship shown in Fig. \ref{Fig:3-2c}, which instead of a quadratic digressive curve, now follows a quadratic progressive curve. 
This nonlinear effect of increasing shim thickness has also been noted by \citet{skavckauskas2017development} for regular to high velocities by test bench experiments and numerical simulations. It implies that the shim thickness has a stronger impact on damping force increase than shim number, due to its direct implication on shim stiffness as opposed to cumulative shim stack stiffness involving contact force pairs at the interfaces between separate shim plates. The physical mechanism and derivation of this cumulative shim stack stiffness is described in more detail in \citet{schickhofer2023universal}.

\subsection{\textbf{Shim diameter}}
\label{Sec:3-3}

\begin{figure}[htbp]
\begin{centering}
\begin{overpic}[width=1.00\textwidth]{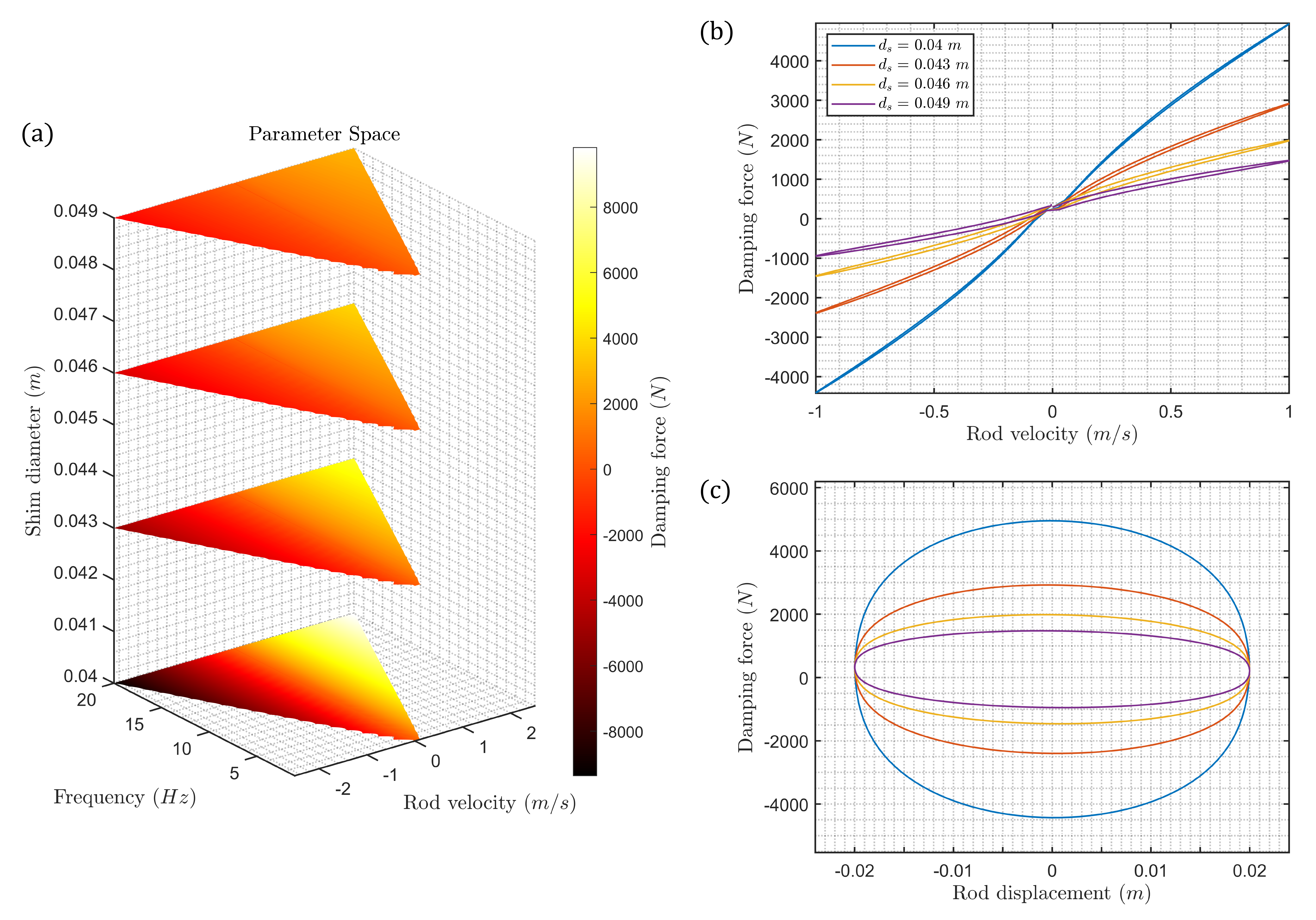}
\end{overpic}
\par
\end{centering}
\caption{Plane sections of the parameter space for damping force as a function of shim diameter, frequency, and rod velocity (a). Damping curves are shown for discrete values of the shim diameter (b),(c).}
\label{Fig:3-3a} 
\end{figure}

\begin{figure}[htbp]
\begin{centering}
\begin{overpic}[width=1.00\textwidth]{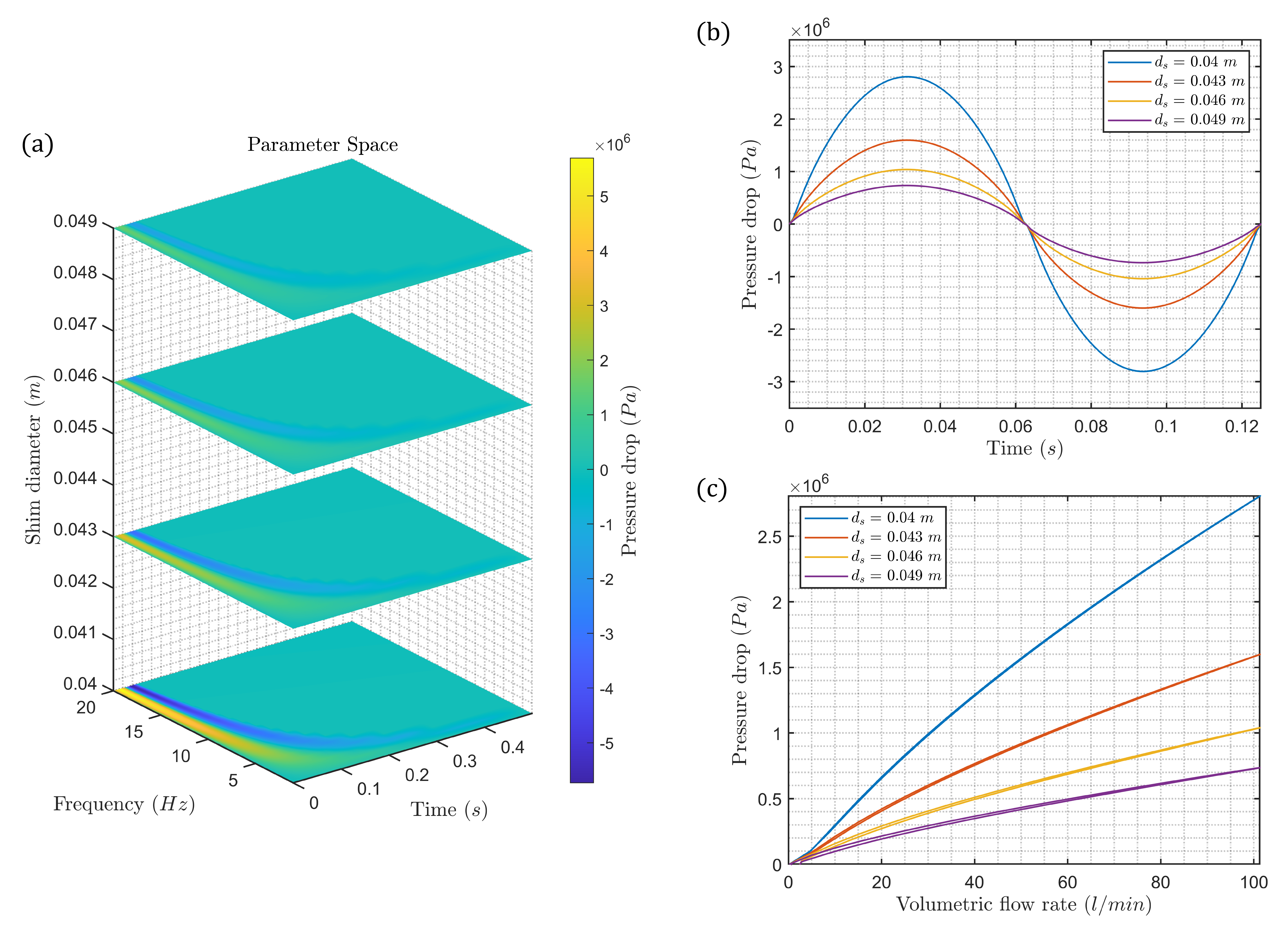}
\end{overpic}
\par
\end{centering}
\caption{Pressure drop as a function of shim diameter, frequency, and time (a). The pressure loss distribution is shown for the entire damping cycle (b) and range of flow rates (c).}
\label{Fig:3-3b} 
\end{figure}

\begin{figure}[htbp]
\begin{centering}
\begin{overpic}[width=1.00\textwidth]{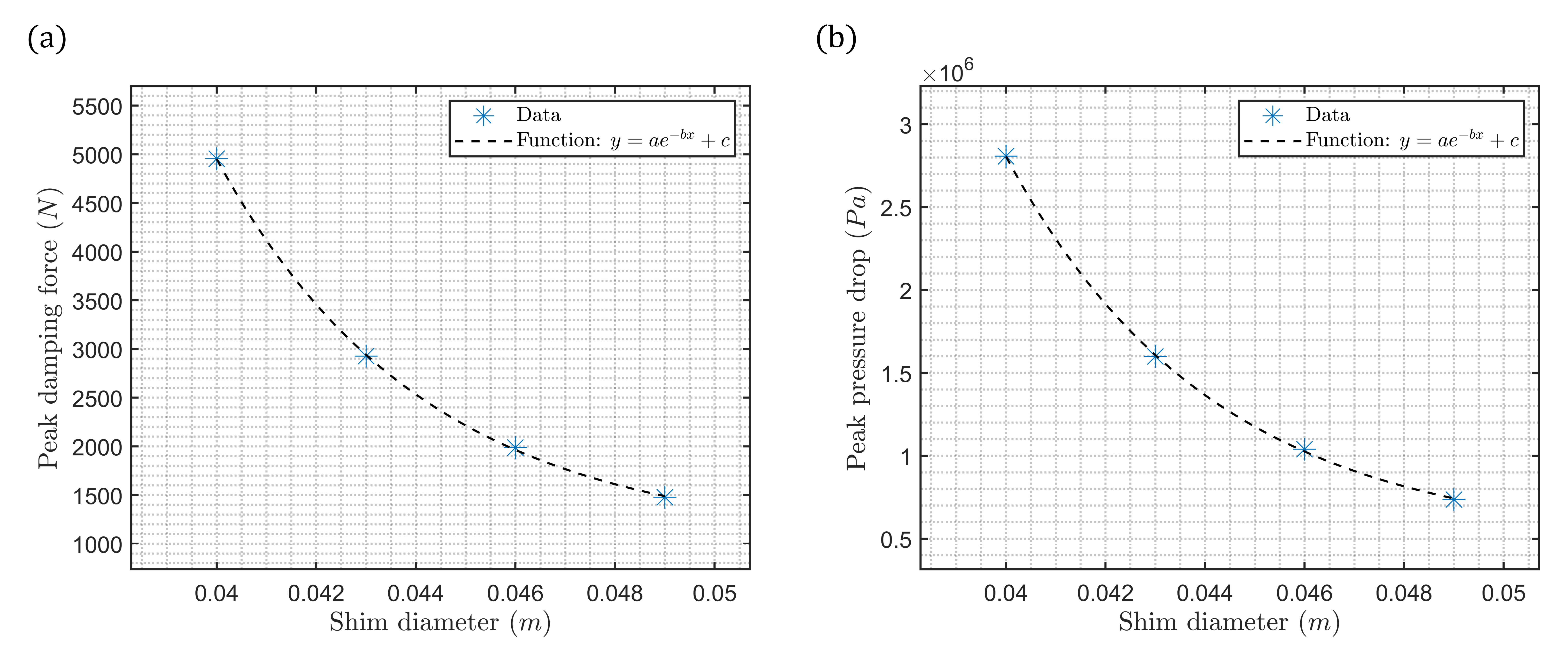}
\end{overpic}
\par
\end{centering}
\caption{Cause-effect relationship between damping force (a), as well as pressure drop (b) and shim diameter. The fitting parameters are $\left[a,b,c\right]=\left[5.95\times10^{7},240.74,1.04\times10^{3}\right]$ with $R^{2}=0.9997$ in (a) and $\left[a,b,c\right]=\left[3.52\times10^{10},240.54,4.75\times10^{5}\right]$ with $R^{2}=0.9998$ in (b).}
\label{Fig:3-3c} 
\end{figure}

As opposed to shim number and thickness, an increase in shim diameter leads to an inverse trend in damping properties, as demonstrated in Fig. \ref{Fig:3-3a}.
Thus, higher shim diameters are typically connected with shim stacks of reduced stiffness.
The nonlinear loss of pressure resulting from shims of larger diameter is visualized in Fig. \ref{Fig:3-3b}(a)-(b), which is particularly apparent for the considered range of flow rates in Fig. \ref{Fig:3-3b}(c).
The cause-effect relationship for damping force and pressure drop follows a steep exponential drop that is significant already at relatively small deviations of only a few millimeters (cf. Fig. \ref{Fig:3-3c}).
This means in practice that the damping potential of a shock absorber in terms of its reaction force and characteristic pressure loss can be generally increased by shims of smaller diameter leading to larger geometric stiffness.

\subsection{\textbf{Bleed orifice area}}
\label{Sec:3-6}

\begin{figure}[htbp]
\begin{centering}
\begin{overpic}[width=1.00\textwidth]{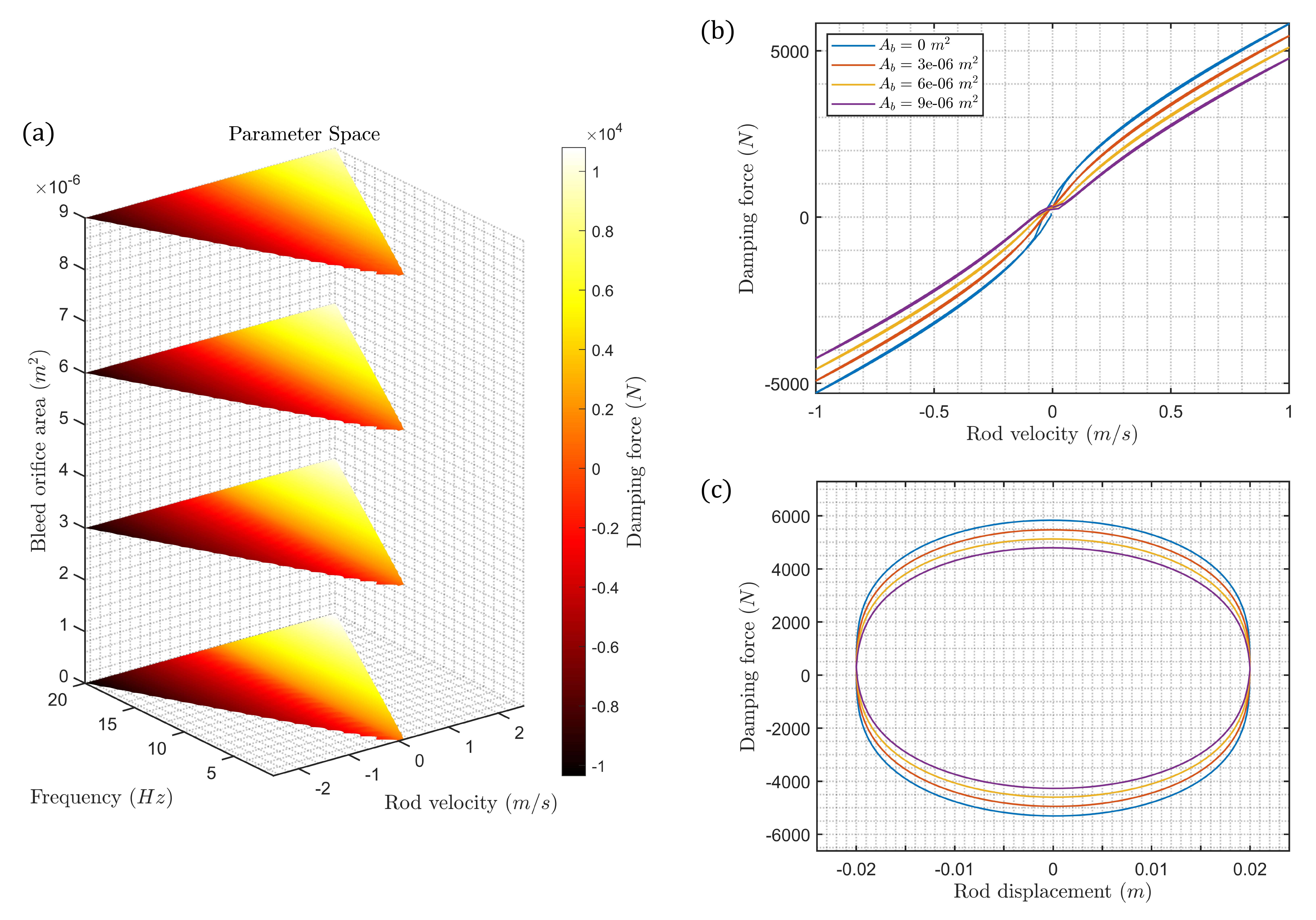}
\end{overpic}
\par
\end{centering}
\caption{Damping force as a function of bleed orifice area, frequency, and rod velocity (a). Damping curves are plotted for rod velocity (b) and displacement (c).}
\label{Fig:3-6a} 
\end{figure}

\begin{figure}[htbp]
\begin{centering}
\begin{overpic}[width=1.00\textwidth]{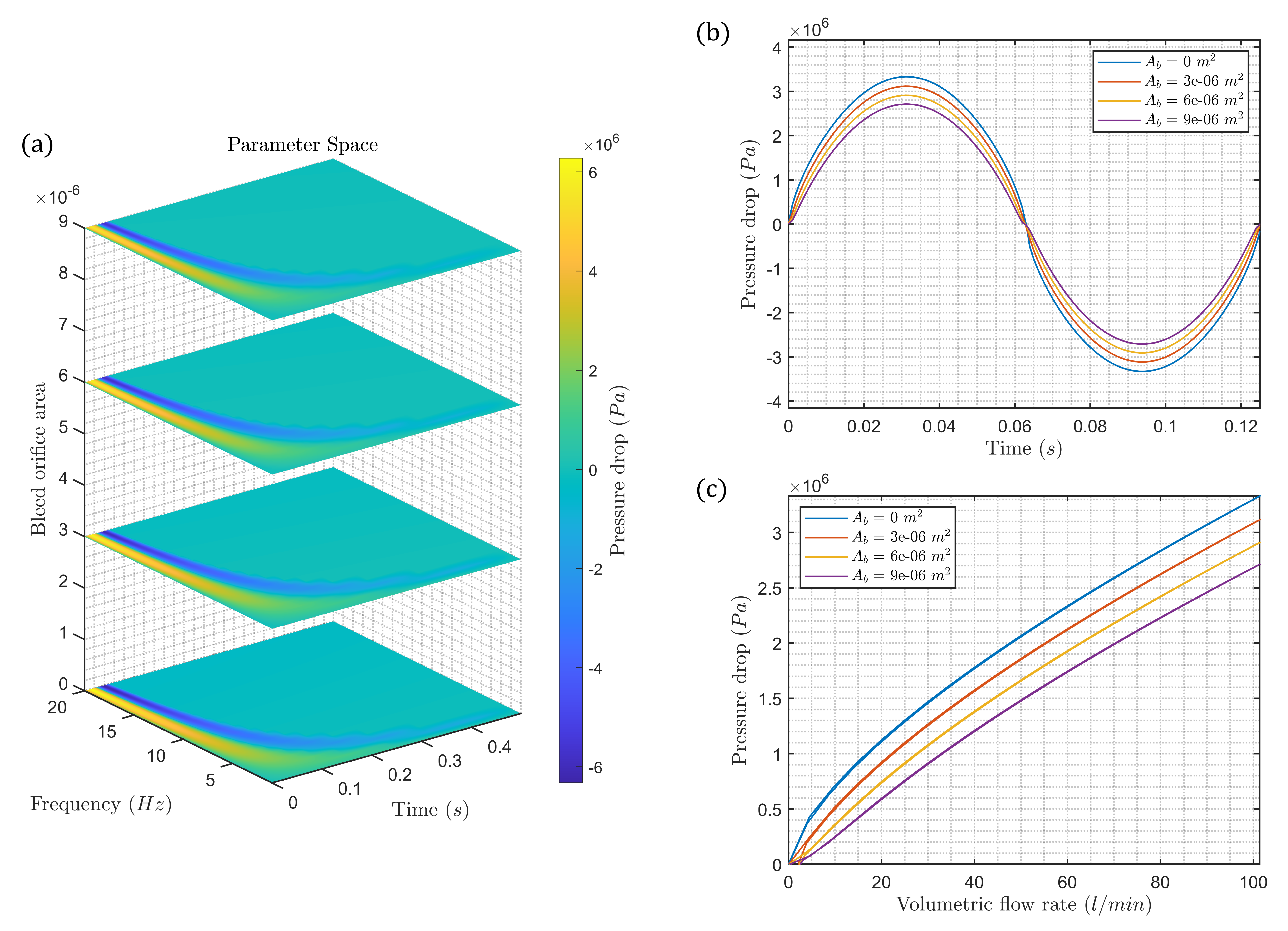}
\end{overpic}
\par
\end{centering}
\caption{Parameter space of pressure drop as a function of bleed orifice area, frequency, and time (a). Time series of pressure loss (b), and its relation to flow rate (c).}
\label{Fig:3-6b} 
\end{figure}

\begin{figure}[htbp]
\begin{centering}
\begin{overpic}[width=1.00\textwidth]{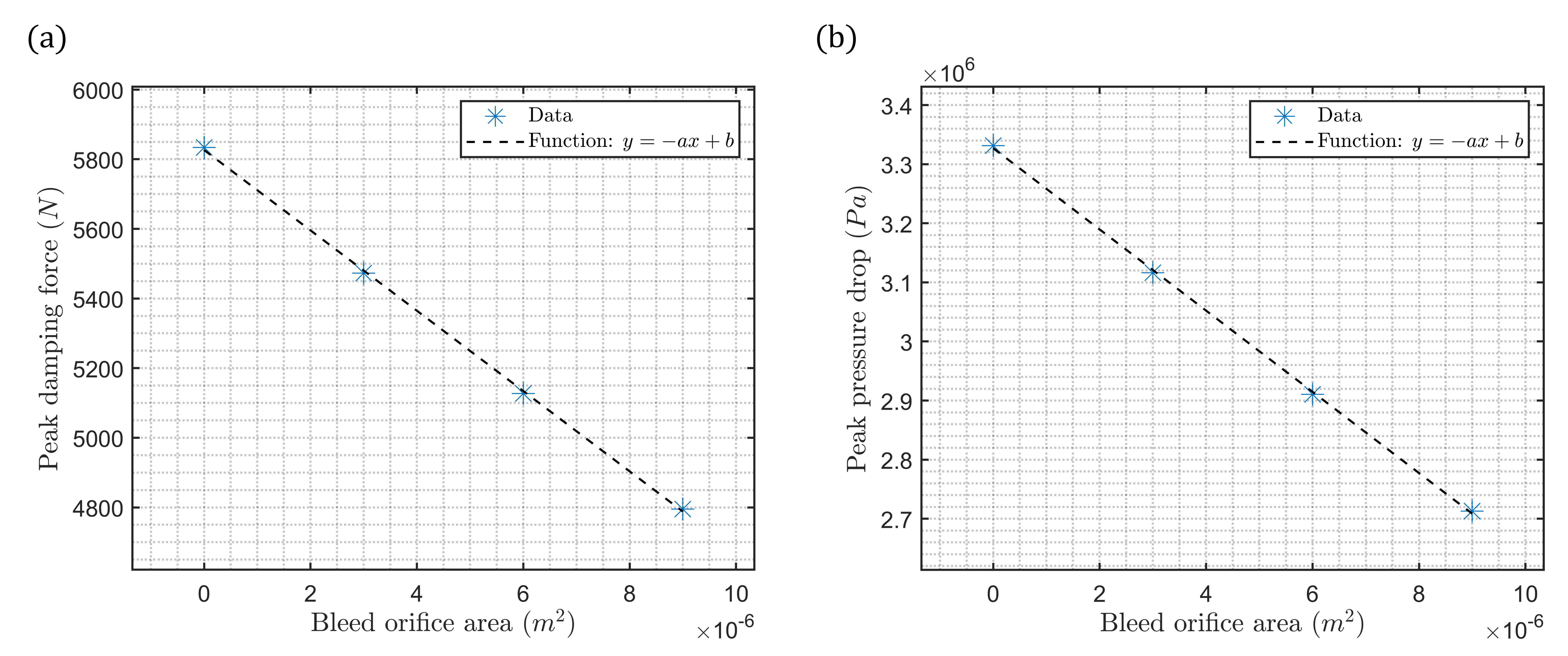}
\end{overpic}
\par
\end{centering}
\caption{Cause-effect relationship between damping force and bleed area (a), as well as pressure drop and bleed area (b). Fitting of the parameters gives $\left[a,b\right]=\left[1.15\times10^{8},5.83\times10^{3}\right]$ with $R^{2}=0.9997$ in (a) and $\left[a,b\right]=\left[6.87\times10^{10},3.33\times10^{6}\right]$ with $R^{2}=0.9998$ in (b).}
\label{Fig:3-6c} 
\end{figure}

One of the most direct and immediate ways of adapting the damping properties of a hydraulic shock absorber is by changing the cross-sectional area of the bleed orifice, a constant constriction located in the centre of the piston that allows for a bleed flow $Q_{b}$, as indicated in Fig. \ref{Fig:2-1}(b). 
This is also how the suspension of motorcycles, bikes, and cars can be quickly adjusted by the user via a small needle blocking parts of the bleed orifice, which allows for modification of the characteristic curves in Fig. \ref{Fig:3-6a}(b)-(c) across the entire rod displacement and velocity range. The advantage of this tuning option is its predictable and linear impact on the shock absorber characteristics, as can be seen in Fig. \ref{Fig:3-6c}. Any changes in the pressure drop as a result of bleed orifice variation are transmitted across the excitation frequency and flow rate range (cf. Fig. \ref{Fig:3-6b}).

\subsection{\textbf{Leakage width}}
\label{Sec:3-7}

\begin{figure}[htbp]
\begin{centering}
\begin{overpic}[width=1.00\textwidth]{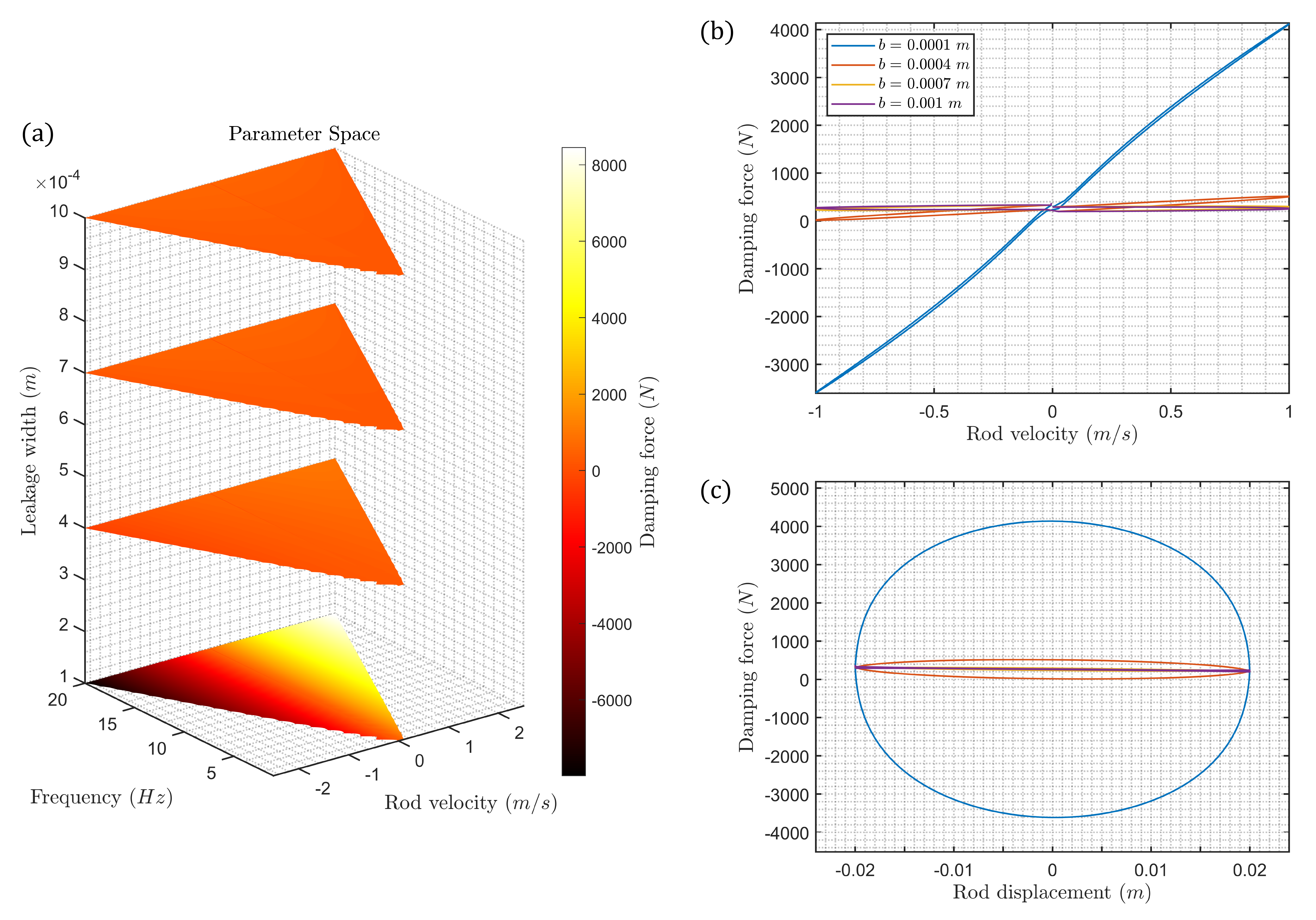}
\end{overpic}
\par
\end{centering}
\caption{Damping force as a function of leakage width, excitation frequency, and rod velocity (a). Characteristic curves of the damping force for various discrete leakage gaps (b),(c).}
\label{Fig:3-7a} 
\end{figure}

\begin{figure}[htbp]
\begin{centering}
\begin{overpic}[width=1.00\textwidth]{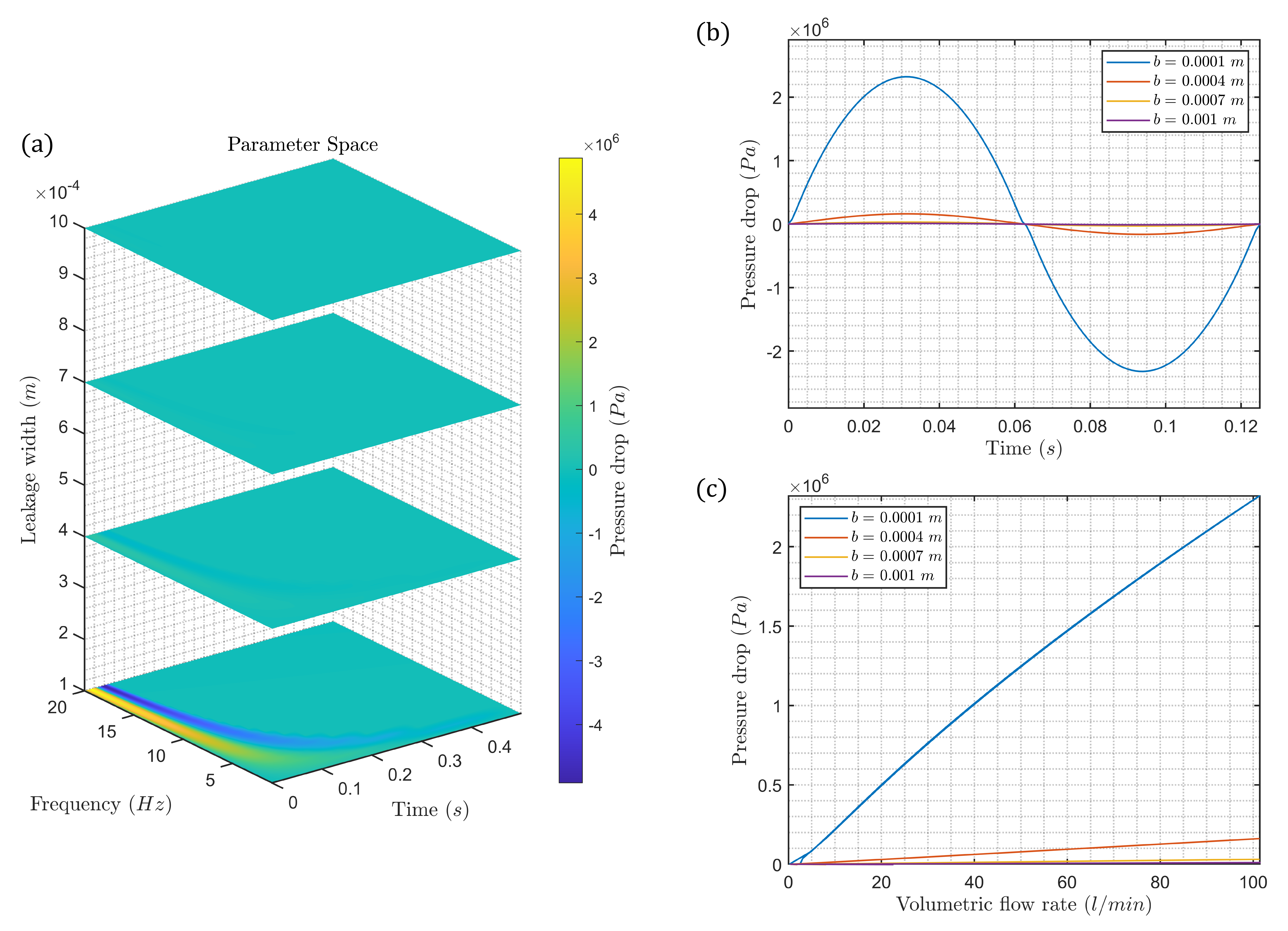}
\end{overpic}
\par
\end{centering}
\caption{Pressure drop as a function of leakage width, frequency, and time (a). Losses of pressure across the piston in time (b) and as a function of volumetric flow rate (c).}
\label{Fig:3-7b} 
\end{figure}

\begin{figure}[htbp]
\begin{centering}
\begin{overpic}[width=1.00\textwidth]{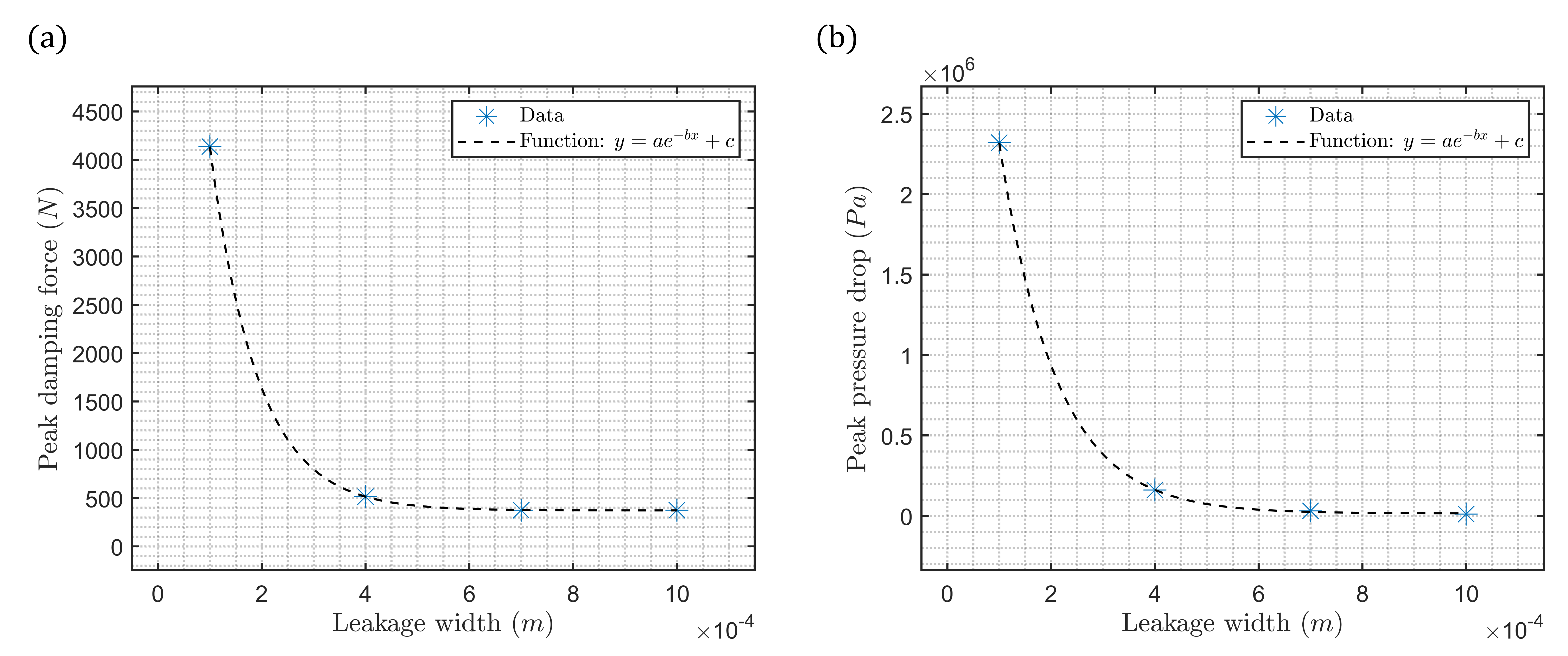}
\end{overpic}
\par
\end{centering}
\caption{Cause-effect relationship of peak damping force (a), as well as peak pressure drop and leakage width (b). Fitting parameters are $\left[a,b,c\right]=\left[1.12\times10^{4},1.09\times10^{4},371.97\right]$ with $R^{2}=0.9999$ in (a) and $\left[a,b,c\right]=\left[5.78\times10^{6},9.19\times10^{3},1.55\times10^{4}\right]$ with $R^{2}=0.9999$ in (b).}
\label{Fig:3-7c} 
\end{figure}

The leakage between the piston walls and the shock absorber compartment, as pictured in Fig. \ref{Fig:2-1}, can cause a considerable loss of damping force. Fig. \ref{Fig:3-7a} indicates that already a small increase in the width of the leakage gap leads to a reduction of damping force of several orders of magnitude, which is essentially triggered by a loss of built-up pressure difference, plotted in Fig. \ref{Fig:3-7b}. Both peak damping force and pressure drop undergo a sharp exponential drop within only $\SI{0.2}{\milli\meter}$, before eventually reaching a plateau (cf. Fig. \ref{Fig:3-7c}).

\subsection{\textbf{Temperature}}
\label{Sec:3-8}

\begin{figure}[htbp]
\begin{centering}
\begin{overpic}[width=1.00\textwidth]{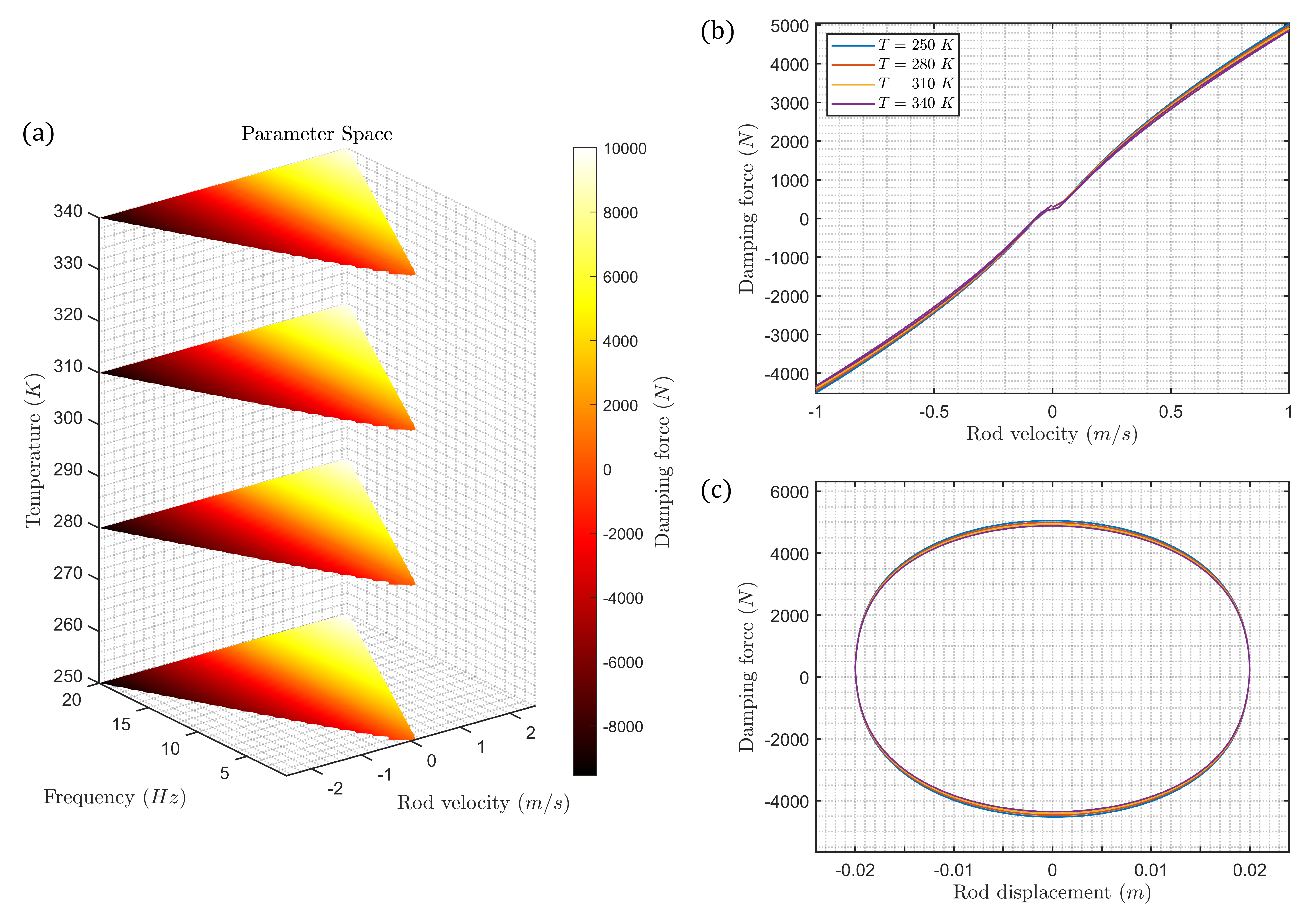}
\end{overpic}
\par
\end{centering}
\caption{Damping force as a function of temperature within a physical range of the shock absorber, frequency, and rod velocity (a). Characteristic curves for rod velocity (b) and displacement (c).}
\label{Fig:3-8a} 
\end{figure}

\begin{figure}[htbp]
\begin{centering}
\begin{overpic}[width=1.00\textwidth]{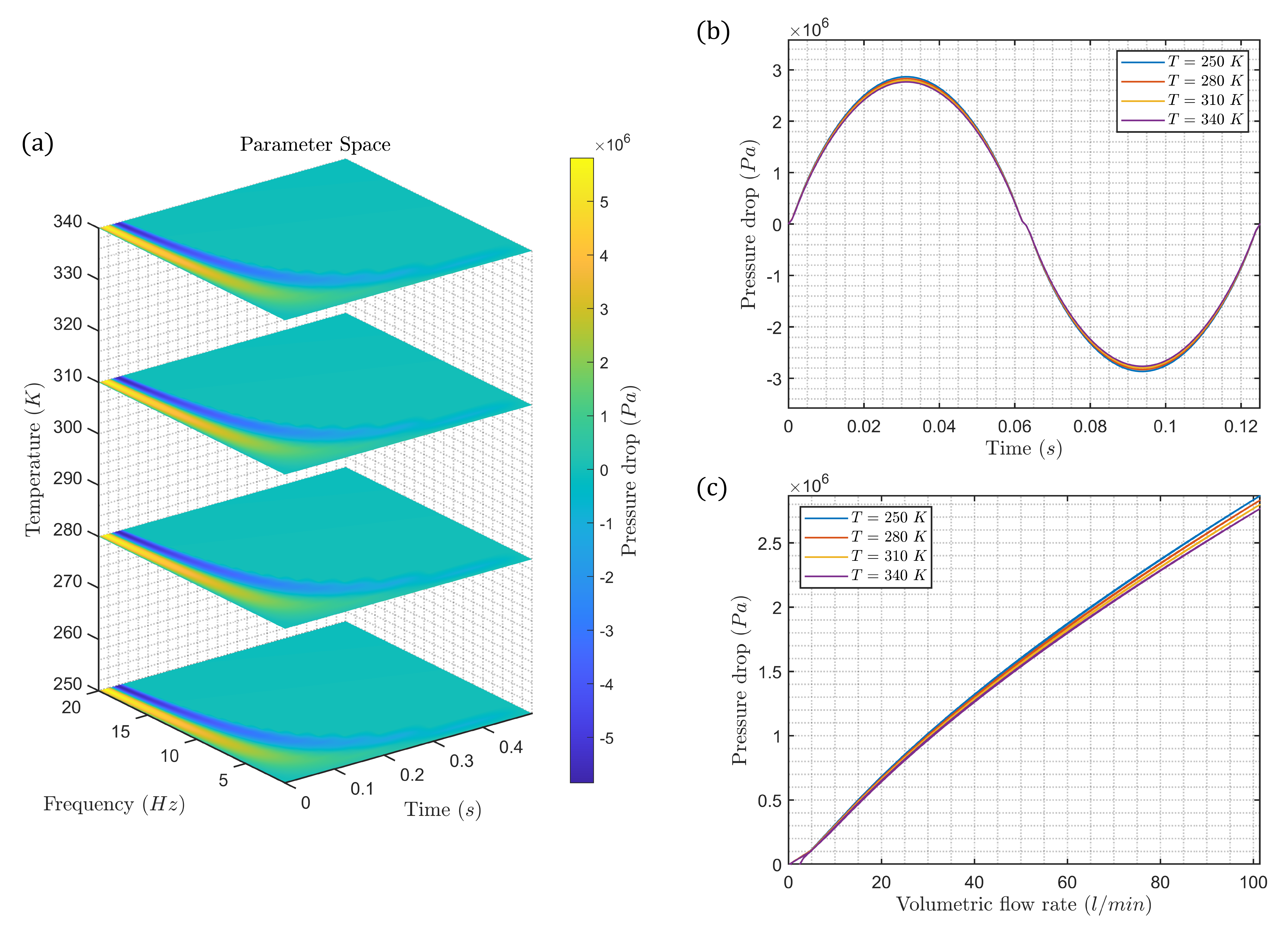}
\end{overpic}
\par
\end{centering}
\caption{Pressure drop as a function of temperature, frequency, and rod velocity (a). Pressure loss characteristics in time (b), and as a function of flow rate (c).}
\label{Fig:3-8b} 
\end{figure}

\begin{figure}[htbp]
\begin{centering}
\begin{overpic}[width=1.00\textwidth]{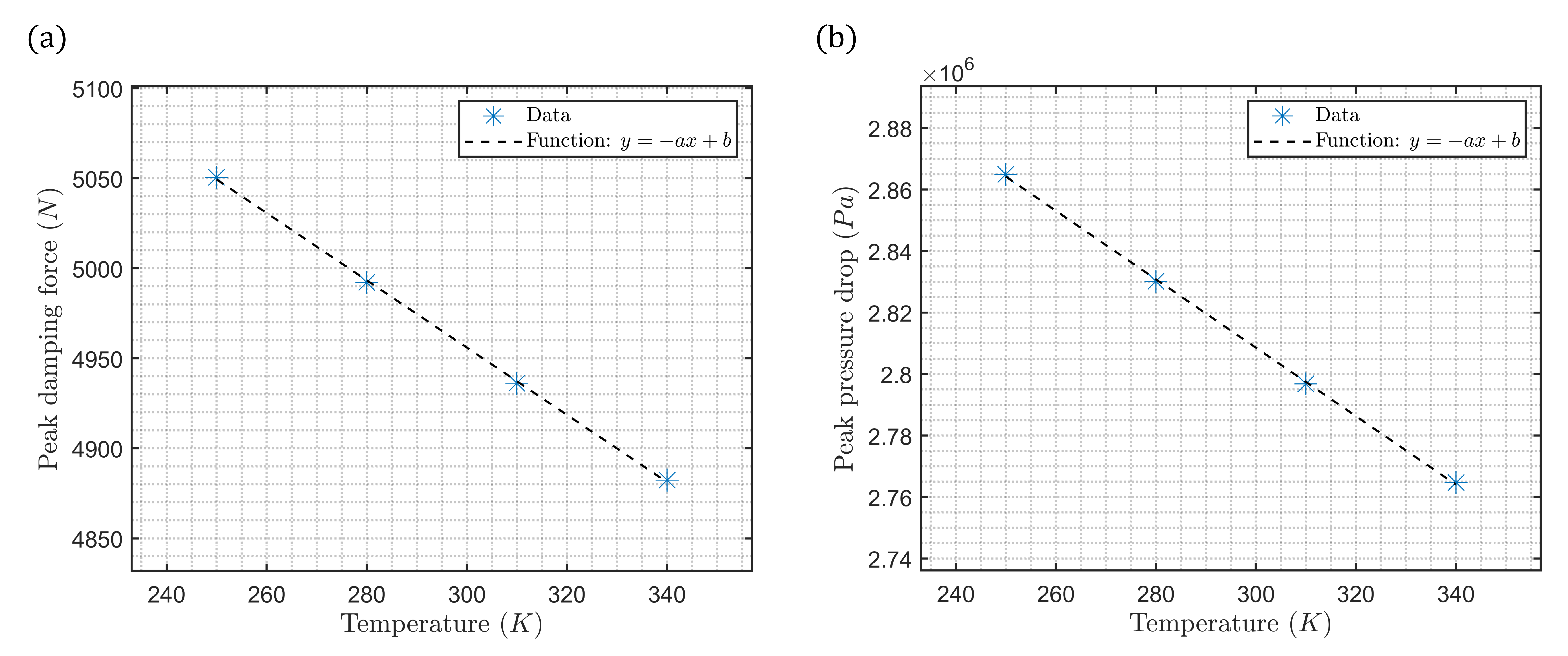}
\end{overpic}
\par
\end{centering}
\caption{Cause-effect relationship between damping force and temperature (a), as well as pressure drop and temperature (b). Fitting parameters are found as $\left[a,b\right]=\left[1.87,5.52\times10^{3}\right]$ with $R^{2}=0.9989$ in (a) and $\left[a,b\right]=\left[1.11\times10^{3},3.14\times10^{6}\right]$ with $R^{2}=0.9985$ in (b).}
\label{Fig:3-8c} 
\end{figure}

Finally, changes in temperature affect various material properties such as density and viscosity, and thus lead to additional unpredictability of the shock absorber damping potential over time, as the damper and its contained mineral oil heats up during prolonged operation and over multiple damping cycles.
However, as demonstrated in Fig. \ref{Fig:3-8a}-\ref{Fig:3-8b}, these effects are generally small and become more pronounced at higher rod velocities and flow rates. Moreover, the cause-effect relationships given in Fig. \ref{Fig:3-8c} show a linear trend.
It must be emphasized here that the underlying model in its current form considers the damper oil to be homogeneous and does not take into account multiphase effects, for instance in the case of dissolved air bubbles. Under such conditions, temperature would most likely have a more pronounced effect on damping properties \citep{duym1997physical}.

\subsection{\textbf{Constriction discharge coefficient}}
\label{Sec:3-4}

\begin{figure}[htbp]
\begin{centering}
\begin{overpic}[width=1.00\textwidth]{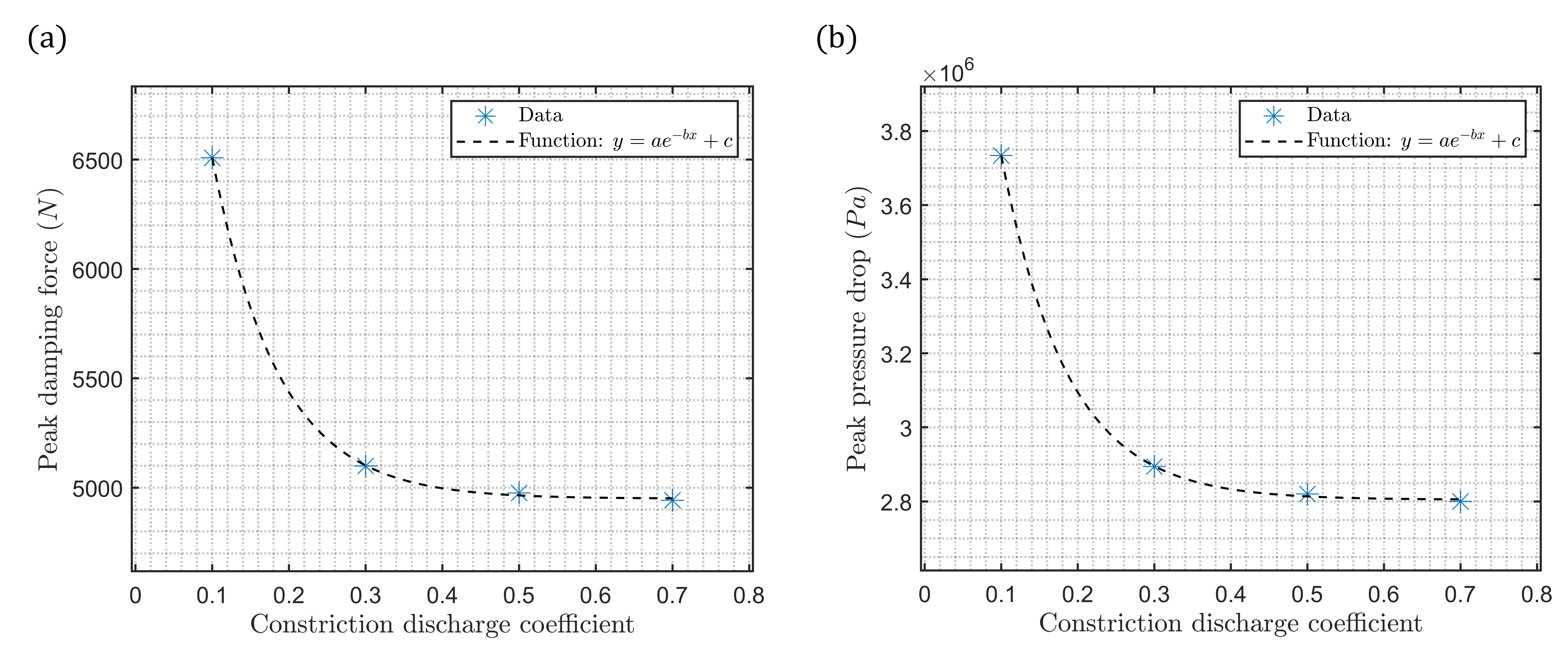}
\end{overpic}
\par
\end{centering}
\caption{Cause-effect relationship of peak damping force (a), as well as peak pressure drop (b) on constriction discharge coefficient. Parameters for the fitting curves are $\left[a,b,c\right]=\left[5.00\times10^{3},11.64,4.95\times10^{3}\right]$ with $R^{2}=0.9996$ in (a) and $\left[a,b,c\right]=\left[2.98\times10^{6},11.65,2.80\times10^{6}\right]$ with $R^{2}=0.9995$ in (b).}
\label{Fig:3-4c} 
\end{figure}

Besides the immediate modifications of a shock absorber via its rebound and compression shim stacks (cf. Sec. \ref{Sec:3-1}-\ref{Sec:3-3}), there remains the possibility of changing geometric properties of the piston and orifices.
One such measure is the discharge coefficient at the entry of the piston channels visible in Fig. \ref{Fig:2-1}(b), referred to here as \emph{constriction discharge coefficient}. Depending on the shape of the entry, it can reach values of $0.6<C_{d,c}<1.0$ for highly cambered and rounded edges, or lower values of $C_{d,c}\leq0.6$ in case of sharp edges and possible shear flow and secondary flow structures, such as recirculation regions \citep{schickhofer2022fluid}.
The effect of the entry piston geometry in the form of its discharge coefficient is significant at high rod velocities \citep{lang1977study}. This is the reason for applying geometric variations with the particular goal of influencing high-frequency damping characteristics. Additionally, reduced flow discharge affects a rapid loss of damping force, as clearly visible in Fig. \ref{Fig:3-4c}. Although \citet{dixon2008shock} notes that the range of typical discharge coefficients of unsteady internal shock absorber flows has been empirically established as $0.6<C_{d}<0.7$, lower values are likely in the aforementioned conditions of a large crossflow component \citep{feseker2018experimental}.

%

\subsection{\textbf{Valve discharge coefficient}}
\label{Sec:3-5}

\begin{figure}[htbp]
\begin{centering}
\begin{overpic}[width=1.00\textwidth]{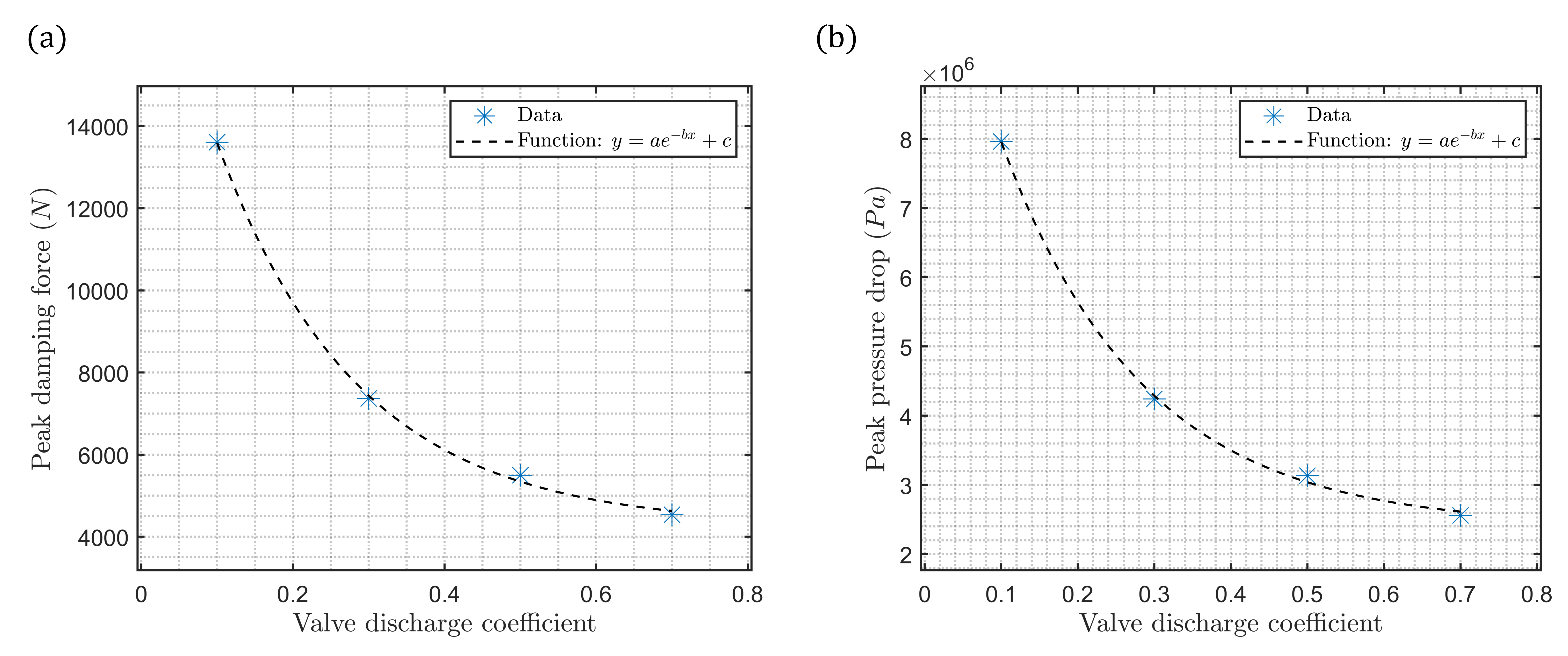}
\end{overpic}
\par
\end{centering}
\caption{Cause-effect relationship of peak values of damping force (a), as well as pressure drop and valve discharge coefficient (b). Fitting parameters are established as $\left[a,b,c\right]=\left[1.60\times10^{4},5.39,4.26\times10^{3}\right]$ with $R^{2}=0.9956$ in (a) and $\left[a,b,c\right]=\left[9.53\times10^{6},5.39,2.39\times10^{6}\right]$ with $R^{2}=0.9959$ in (b).}
\label{Fig:3-5c} 
\end{figure}

Congruent with the entry discharge coefficient of the shock absorber piston, there is also an exit quantity defining the respective geometric shape referred to as \emph{valve discharge coefficient}. The exit geometry of the piston is crucial for the cross section of the jet flow acting on the shim stack and therefore its pressure area. 
Unlike the constriction discharge coefficient, which is acting as a preset high-speed damping before the main pressure loss mechanism through frictional effects and elastic deformation of the shim stack, the valve discharge coefficient affects damping across the entire velocity range, as demonstrated in Fig. \ref{Fig:3-5c}.

%

\section{\textbf{Conclusions}}
\label{Sec:4}

The analytical model presented in Sec. \ref{Sec:2} can capture the dynamic behaviour of shock absorbers and allows for an in-depth study of its functional dependence on model parameters. Thus, an extensive database comprising maps of multi-dimensional parameter spaces is created, which shows the immediate impact of changes of the baseline setup on its damping performance (cf. Sec. \ref{Sec:3}). 

In order to obtain cause-effect relationships in the form of diagrams showing the dependence of the target quantities (damping force, pressure drop) on parameter changes (e.g. shim properties, bleed orifice area, temperature), nonlinear curve fitting is performed with a goodness-of-fit consistently above $R^{2}=0.99$.

While some of the investigated parameters, such as the bleed orifice area or temperature have a mostly linear effect on pressure losses and damping curves, the variation of others causes a severe change in the behaviour of the system:
\begin{itemize}
\item Shim number and thickness both show a quadratic relationship to damping force. However, while the damping curves eventually flatten with a large number of shims due to their digressive trend, the cause-effect relationship for increasing shim thickness follows a progressive trend. This is in agreement with experimental and numerical results by \citet{skavckauskas2017development}, who also pointed out that it takes a number of $n_{s}\sim(t_{s,2}/t_{s,1})$ shims of smaller thickness $t_{s,1}$ to account for the damping effect of a larger shim thickness $t_{s,2}$.
\item Furthermore, by choosing a larger shim diameter, one can reach an exponentially lower damping force. More precisely, a $\SI{25}{\percent}$ increase in diameter causes a drop in damping force of almost $\SI{70}{\percent}$.
\item The variation of the constriction inlet discharge coefficient demonstrates the impact of this geometric parameter on the high-speed damping behaviour. It affects the pressure loss and damping characteristics almost exclusively at low values and high stroking frequencies, which makes it the ideal property for shock absorber designers to adapt, if this range is of interest for further tuning. The valve outlet discharge coefficient on the other hand shows a more gradual impact across a large range of volumetric flow rates, but its peak damping increase at smaller values is also of an exponential nature. 
\item Linear relationships of bleed orifice variation and maximum damping properties are explained by its linear proportionality with bleed flow rates. The results here also confirm the linear effect of bleed orifices on damping force observed by \citet{talbott2002experimentally} with their experimentally validated model.
\item Finally, a total loss of damping by the shock absorber system occurs already at small increments of the leakage width, which quantifies the gap between the piston and the damper compartment.
\end{itemize}

The findings above are also crucial for the derivation of further cause-effect curves based on computed fitting parameters. Thus, a clear picture emerges of the overall dynamical behaviour of the system as it undergoes tuning and changes in its parameter settings. Most importantly, it provides the first comprehensive study on the sensitivity of the most common shock absorber architecture towards setup modifications.
\bigskip

\noindent
\textbf{Acknowledgements} \\
The work has been performed under the Project HPC-EUROPA (INFRAIA-2016-1-730897), with the support of the European Commission for Research and Innovation under the H2020 Programme (Grant: HPC17ZLPYQ). In particular, the authors gratefully acknowledge the support of the Department of Mathematical Sciences at the University of Essex and the computer resources and technical support provided by the Irish Centre for High-End Computing (ICHEC).
\bigskip

\noindent
\textbf{Conflict of interest statement} \\
The authors have nothing to disclose that would have biased this work.


\appendix

\section{Baseline system properties}
\label{App:A}

\setcounter{table}{0}
\setcounter{equation}{0}

Tables \ref{Tab:A-1}-\ref{Tab:A-3} list the basic properties for the shock absorber system used for the computation of the results presented in Sec. \ref{Sec:3}.
These values are based on a realistic suspension setup as applied within racing cars and offroad motorcycles.
Wherever other values are used for the purpose of parametric studies, this is explicitly stated in the text.
For the shim stacks, the following setup of circular annular shims is used:
\begin{itemize}
\item The compression stack consists of shims with radii \\ 
$\left[a_{k}\right]=\left[12,13,14,15,16,18,19,20,21,22,22,22,22,16,22,22,22,22,22.5\right]\times10^{-3}\,\SI{}{\meter}$ and thicknesses $\left[t_{k}\right]=[0.20,0.25,0.25,0.20,0.20,0.20,0.20,0.20,0.25,0.20,$ \linebreak $0.20,0.20,0.20,0.10,0.20,0.20,0.20,0.20,0.20]\times10^{-3}\,\SI{}{\meter}$ with a clamp radius of $a_{c}=11\times10^{-3}\,\SI{}{\meter}$.
\item The rebound stack contains shims with radii $\left[a_{k}\right]=\left[13,14,15,16,17,18,12,14,20,20,20,20,20\right]\times10^{-3}\,\SI{}{\meter}$ and thicknesses $\left[t_{k}\right]=[0.30,0.30,0.30,0.30,0.25,0.25,0.10,0.10,0.15,0.20,0.20,0.20,0.20]\times10^{-3}\,\SI{}{\meter}$ with a clamp radius of $a_{c}=11.5\times10^{-3}\,\SI{}{\meter}$.
\end{itemize}

\begin{table}[htbp]
\caption{Properties of the shock absorber pistons and compartment.}
\centering
\begin{tabular}{@{}lllll@{}}
\toprule
Parameter & Symbol & Value & Unit \\
\midrule
Reservoir pressure & $p_{g}$ & $10^{6}$ & $\SI{}{\pascal}$ \\
Reservoir length & $L_{g}$ & $70.11\times10^{-3}$ & $\SI{}{\meter}$ \\
Rebound chamber length & $L_{1}$ & $23.90\times10^{-3}$ & $\SI{}{\meter}$ \\
Compression chamber length & $L_{2}$ & $157.18\times10^{-3}$ & $\SI{}{\meter}$ \\
Gas piston diameter & $d_{g}$ & $59.78\times10^{-3}$ & $\SI{}{\meter}$ \\
Main piston diameter & $d_{piston}$ & $49.60\times10^{-3}$ & $\SI{}{\meter}$ \\
Rod diameter & $d_{rod}$ & $17.97\times10^{-3}$ & $\SI{}{\meter}$ \\
Gas piston mass & $m_{g}$ & $0.1955$ & $\SI{}{\kilogram}$ \\
Main piston and rod mass & $m_{tot}$ & $1.0980$ & $\SI{}{\kilogram}$ \\
Cylinder radius & $r_{d}$ & $25\times10^{-3}$ & $\SI{}{\meter}$ \\
Wall thickness & $t_{d}$ & $2.5\times10^{-3}$ & $\SI{}{\meter}$ \\
Stainless steel Young's modulus & $E_{d}$ & $200\times10^{9}$ & $\SI{}{\pascal}$ \\
Seal friction & $F_{f}$ & $22.25$ & $\SI{}{\newton}$ \\
Rebound discharge coefficients & $[C_{d,c},C_{d,v}]$ & $[0.34,0.5]$ & $\SI{}{-}$ \\
Compression discharge coefficients & $[C_{d,c},C_{d,v}]$ & $[0.5,0.7]$ & $\SI{}{-}$ \\
Bleed orifice discharge coefficient & $C_{d,b}$ & 0.6 & $\SI{}{-}$ \\
\bottomrule
\end{tabular}
\label{Tab:A-1}
\end{table}

\begin{table}[htbp]
\caption{Properties of the shim stacks.}
\centering
\begin{tabular}{@{}lllll@{}}
\toprule
Parameter & Symbol & Value & Unit \\
\midrule
Shim steel Poisson's ratio & $\nu$ & $0.305$ & $\SI{}{-}$ \\
Shim steel Young's modulus & $E$ & $210\times10^{9}$ & $\SI{}{\pascal}$ \\
\bottomrule
\end{tabular}
\label{Tab:A-2}
\end{table}

\begin{table}[htbp]
\caption{Properties of the damper mineral oil.}
\centering
\begin{tabular}{@{}lllll@{}}
\toprule
Parameter & Symbol & Value & Unit \\
\midrule
Reference density & $\rho_{0}$ & $830$ & $\SI{}{\kilogram\per\cubic\meter}$ \\
Reference dynamic viscosity & $\mu_{0}$ & $0.01$ & $\SI{}{\kilogram\per\meter\per\s}$ \\
Reference compressibility & $\beta_{0}$ & $6.6\times10^{-10}$ & $\SI{}{\kilogram\per\cubic\meter}$ \\
Reference temperature & $T_{0}$ & $288$ & $\SI{}{\kelvin}$ \\
Temperature sensitivity coefficient & $C$ & $3717$ & $\SI{}{\kelvin}$ \\
Thermal expansion coefficient & $\alpha_{T}$ & $0.001$ & $\SI{}{\per\kelvin}$ \\
\bottomrule
\end{tabular}
\label{Tab:A-3}
\end{table}

\newpage

\end{document}